\documentclass[a4paper, 11pt]{article}
\pdfoutput=1
\usepackage{jcappub}
\usepackage{graphicx}
\usepackage{color}
\usepackage{times}
\usepackage[T1]{fontenc}
\usepackage{bm}
\usepackage{ulem}
\usepackage{multirow}
\usepackage{float}
\usepackage{url}
\usepackage{natbib}
\usepackage{mathrsfs}
\usepackage{physics}
\usepackage{comment}
\usepackage[table,xcdraw]{xcolor}
\usepackage{booktabs,makecell}
\usepackage{comment}
\usepackage[colorlinks=true,citecolor=blue,urlcolor=blue,linkcolor=blue]{hyperref}
\newcommand{\orcid}[1]{\href{https://orcid.org/#1}{\includegraphics[keepaspectratio,width=0.8em]{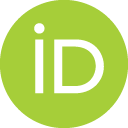}}}

\title{Probing the impact of Delta-Baryons on Nuclear Matter and Non-Radial Oscillations in Neutron Stars}
\author[a]{Probit J Kalita,\orcid{0000-0002-5196-3603}}
\author[a]{Pinku Routaray,\orcid{0000-0002-6746-7719}}
\author[a]{Sayantan Ghosh,\orcid{0000-0001-8276-1935}}
\author[a]{Bharat Kumar,\orcid{0000-0003-1964-057X}}
\author[b,c]{and B. K. Agrawal}
\affiliation[a]{Department of Physics and Astronomy, National Institute of Technology, Rourkela 769008, India}
\affiliation[b]{Saha Institute of Nuclear Physics, Kolkata, 700064, India}
\affiliation[c]{Homi Bhabha National Institute, Training School Complex, Anushakti Nagar, Mumbai 400094, India}
\emailAdd{probitjkalita@disroot.org}
\emailAdd{routaraypinku@gmail.com}
\emailAdd{sayantanghosh1999@gmail.com}
\emailAdd{kumarbh@nitrkl.ac.in}
\abstract{
The presence of heavy baryons, such as $\Delta$-baryons and hyperons can significantly impact various properties of  Neutron Stars (NSs), like oscillation frequencies, dimensionless tidal deformability, mass, and radii. We explored these effects within the Density-Dependent Relativistic Mean Field formalism. Our analysis considered $\Delta$-admixed NS matter in both hypernuclear and hyperon-free scenarios, providing insights into particle compositions and their effects on NS properties. Our study of non-radial $f$-mode oscillations revealed a distinct increase in frequency due to the additional baryons. The degree of increase was significantly influenced by the meson-baryon coupling strengths. Notably, the coupling between $\Delta$-resonances and $\sigma$-mesons played a highly influential role. In some cases, it led to an approximately 20\% increase in the $f$-mode oscillation frequency of canonical NSs. These couplings also affect other bulk properties of NSs, including mass, radii, and dimensionless tidal deformability ($\Lambda$). Comparing our results with available observational data from pulsars (NICER) and gravitational waves (LIGO-VIRGO collaboration), we found strong agreement, particularly concerning $\Lambda$.
}
\begin{document}
\maketitle
\flushbottom
\section{Introduction}
\label{sec:Introduction}

Neutron Stars come to be when massive stars reach the end of their life journey as core-collapse supernovae. This transformation sets the stage for a variety of events that trigger oscillations within the star. These oscillations possess sufficient energy to be picked up by instruments designed to detect gravitational waves. These initiating events could be linked to the star's magnetic configuration, dynamic instabilities, accumulation of matter, and fractures in its outer layer~\cite{tsang_resonant_2012, franco_quaking_2000, hinderer_effects_2016, chirenti_gravitational_2017}. Kip Thorne pioneered the study of these disturbances within massive stars using the principles of general relativity~\cite{thorne_non-radial_1967, thorne_nonradial_1969, thorne_nonradial_1969-1, price_non-radial_1969}. Substantial efforts have been invested in extending the basic concepts of oscillation theory from Newtonian physics to the more intricate framework of general relativity. These extensions aim to determine the frequencies at which oscillations occur and quantify the energy emitted in the form of gravitational waves~\cite{zhao_universal_2022, zhao_quasinormal_2022, sotani_universal_2021}.

The exploration of these oscillation frequencies involves solving equations that describe fluid perturbations alongside equations that govern how matter and spacetime curvature interact in the presence of strong gravitational forces~\cite{tolman_static_1939, oppenheimer_massive_1939, cowling_non-radial_1941, sotani_signatures_2011, das_impacts_2021}. These oscillations are categorized into two primary types: radial and non-radial, both of which are subjects of active research. Radial oscillations involve expansions and contractions akin to a pulsating motion that helps maintain the star's spherical shape~\cite{panotopoulos_radial_2020, rather_radial_2023, routaray_radial_2023, sen_radial_2023,pinku_gw-posterior}. In contrast, non-radial oscillations manifest as asymmetric vibrations centered around the star's core are guided by a restoring force that brings the star back to its equilibrium state~\cite{zhao_universal_2022, sotani_universal_2021, zhao_quasinormal_2022, mohanty_impact_2023,bikram_fmode_2021,bikram_GRfmode_2021,athul-fmode_2022,routaray_investigating_2023, routaray_dark_2023}. Non-radial oscillations can manifest in various modes, denoted as $f$, $p$, $g$, $r$, and $w$-modes, although not all of them contribute to the emission of gravitational waves. These modes gradually lose energy and are referred to as quasi-normal modes. The frequencies of these oscillations are significantly influenced by the internal characteristics of the NS, making them valuable tools for probing its interior through the field of asteroseismology. This approach has already provided insights into the properties of the NS's outer layer~\cite{gearheart_upper_2011, sotani_probing_2012, sotani_effect_2013, sotani_possible_2013, sotani_possible_2016, sotani_probing_2017, sotani_constraints_2018, sotani_astrophysical_2019}. NSs hold promise for asteroseismological study via gravitational waves, with expectations that the observation of gravitational waves generated by these oscillations will enable the determination of key properties such as mass, radius, and equation of state (EoS)~\cite{andersson_gravitational_1996, andersson_towards_1998, sotani_density_2001, passamonti_towards_2012, doneva_gravitational_2013}. Among the diverse oscillation modes, the fundamental ($f$) mode stands out as an acoustic oscillation intricately tied to the star's average density ($M/R^3$)~\cite{andersson_gravitational_1996, andersson_towards_1998, sotani_gravitational_2020, sotani_estimating_2020}.

The particle composition in the interior of NSs has been extensively studied since Landau, Baade, and Zwicky first proposed the concept of NSs~\cite{landau1932theory, baade_cosmic_1934}. Over the years, significant work has been conducted in this area, and it has now become conventional to consider the presence of the spin-1/2 baryons octet, also known as hyperons, in the core of NSs~\cite{glendenning_hyperon_1982, prakash_quark-hadron_1995, baldo_hyperon_2000, oertel_hyperons_2015, vidana_hyperons_2016, marquez_phase_2017, roark_hyperons_2019, stone_equation_2021, sedrakian_confronting_2020, menezes_neutron_2021, motta_role_2022}. Additionally, recent studies have also explored the existence of other heavy baryons like the $\Delta$-particles~\cite{marquez_delta_2022, issifu_exotic_2023, li_implications_2019, malfatti_delta_2020, li_rapidly_2020, thapa_equation_2020, backes_effects_2021, thapa_massive_2021, dexheimer_delta_2021, raduta_equations_2022, marczenko_chiral_2022}. These heavy baryons play a crucial role in satisfying the observational constraints on NSs, which have been set by studying massive NSs~\cite{demorest_2010, arzoumanian_nanograv_2018, fonseca_nanograv_2016, ozel_massive_2010}, analyzing the NICER data obtained from various pulsars~\cite{riley_nicer_2021, fonseca_refined_2021, riley_nicer_2019, miller_psr_2019}, and examining gravitational wave data from the LIGO-VIRGO collaboration~\cite{abbott_gw190814_2020, abbott_gw170817_2017}. Among these constraints, special attention is given to the dimensionless tidal deformability $\pqty{\Lambda}$ of the binary NS merger event GW170817, where the reported value was found to be below 720 within the 90\% confidence interval for the canonical NS mass of $1.4M_\odot$~\cite{ligo_virgo_properties_2019}. Achieving such a low value of $\Lambda$ requires a "softening" of the NS matter's EoS (table 1 in \cite{athul-fmode_2022} shows the canonical tidal deformability obtained from nucleonic RMF models like NL3 which are much higher than that observed). This softening can be achieved by including heavier particles such as hyperons~\cite{schaffner_hyperon-rich_1996, wu_strange_2011}, $\Delta$-baryons~\cite{sun_strangeness_2019, maslov_hyperons_2017, kolomeitsev_delta_2017, sedrakian_delta-resonances_2022, drago_early_2014, glendenning_neutron_1985, dexheimer_delta_2021, marquez_delta_2022, schurhoff_neutron_2010}, or (anti)kaons~\cite{thapa_dense_2020, maruyama_finite_2006, brown_strangeness_2006, shao_influence_2010, char_massive_2014} in the matter composition. However, the presence of these particles introduces its own challenges. Hyperons have been found to have a significant impact on NSs, as their nucleation introduces new degrees of freedom in the system which leads to considerable softening of the EoS~\cite{burgio_hyperon_2011, lonardoni_hyperon_2015, bombaci_hyperon_2017}. While this softening is crucial to meet the observed upper bound on $\Lambda$, it also causes the maximum mass configuration that NSs can attain to drop below the observed massive NSs with a mass of $2M_\odot$. This discrepancy is commonly referred to as the "hyperon puzzle". Additionally, owing to their masses lying in a similar range as the hyperons, it should be reasonable to include $\Delta$-baryons into the composition as well, and we can expect them to appear in the NS matter at a similar density range as hyperons~\cite{drago_can_2014, kolomeitsev_delta_2017, raduta_-admixed_2021, takeda_catalysis_2018}. While early works on the topic had ruled out the possibility of the presence of $\Delta$-baryons within NSs~\cite{glendenning_neutron_1985, glendenning_reconciliation_1991}, later works have shown that their presence inside NSs is actually possible given that the $\Delta$-baryon's coupling parameters are properly constrained via available experimental measurements~\cite{li_implications_2019, li_competition_2018, ribes_interplay_2019, xiang_ensuremathdelta_2003, schurhoff_neutron_2010, lavagno_hot_2010, drago_can_2014, drago_early_2014, cai_critical_2015, kolomeitsev_delta_2017, raduta_-admixed_2021}. Similar to hyperons, adding the $\Delta$-baryons also leads to softening of the EoS thereby further decreasing the maximum mass that the NS can attain~\cite{drago_early_2014}.

This calls for the need of some mechanism that can lead to EoSs that are soft enough at the intermediate density range to satisfy the tidal deformability constraints while being stiff enough to result in mass-radius relations that satisfy the observations from massive NSs. Different approaches have been taken with this regard, including but not limited to, adding a repulsive 3-body force~\cite{lonardoni_hyperon_2015}, addition of repulsive interaction between hyperons via the $\phi$ meson~\cite{bhuyan_attribute_2017, schaffner_hyperon-rich_1996, weissenborn_hyperons_2012}, a $\sigma$-cut scheme that aims to keep the EoS stiff at high densities~\cite{ma_kaon_2022, patra_effect_2022, maslov_making_2015, zhang_massive_2018}, and density-dependent coupling constants~\cite{issifu_exotic_2023, marquez_delta_2022, malik_equation--state_2021, sun_neutron_2008, char_massive_2014, thapa_massive_2021, banik_density_2002, char_generalised_2023, colucci_equation_2013, backes_effects_2021}.

The approach adopted in this work to attempt to solve the EoS problem is to use the DD-MEX model~\cite{taninah_parametric_2020} to study the NS matter by including hyperons and $\Delta$-resonance within the framework of the density-dependent relativistic mean field (DDRMF) theory. We also investigate their effects on the various macroscopic properties of NSs, including the dimensionless tidal deformability ($\Lambda$) and the non-radial $f$-mode oscillations. Radial oscillations in NSs for different matter compositions has been an active area of study~\cite{kokkotas_radial_2001, li_oscillation_2022, sen_radial_2023, sagun_asteroseismology_2020, panotopoulos_radial_2018, routaray_radial_2023} with the matter composition being recently extended to include $\Delta$-resonances as well~\cite{rather_radial_2023}. Through this work we are proceeding further by studying, for the first time, non-radial $f$-mode oscillations in NSs with $\Delta$-admixed hypernuclear as well as hyperon-free matter.

We have structured this paper as follows. We first present the theoretical formalism on which we have based our calculations. We follow it up by studying the effects of $\Delta$-baryons and hyperons on NSs with density-dependent couplings. Finally, based on the results obtained we provide some conclusions.

\section{DDRMF Lagrangian and Equation of State}\label{subsec:EOS}
In our study, we use the density-dependent relativistic mean-field (DDRMF) formalism to describe the NS composition. Specifically, we consider that the high density inside the core of a NS facilitates the presence of nucleons (neutrons and protons), hyperons ($\Lambda$, $\Sigma^{+,0,-}$, $\Xi^{0,-}$) and delta baryons ($\Delta^{++,+,0,-}$), with the inter-baryon strong force being mediated by three types of mesons ($\sigma$, $\omega$ and $\rho$). The Lagrangian density resulting from this model is given by~\cite{marquez_delta_2022, issifu_exotic_2023, chen_building_2014},
\begin{align}
    \mathcal{L} = & \sum_{b \in N,H} \bar{\Psi}_b \bqty{\gamma_\mu \pqty{\iota \partial^\mu - g_{\omega b} \omega^\mu - \frac{g_{\rho b}}{2} \va{\tau} \vdot \va{\rho}^\mu } - \pqty{m_b - g_{\sigma b} \sigma}} \Psi_b + \sum_l \bar{\Psi}_l \pqty{\iota \gamma_\mu \partial^\mu - m_l} \Psi_l \nonumber                  \\
                  & + \sum_d \bar{\Psi}_d \bqty{\gamma_\mu \pqty{\iota \partial^\mu - g_{\omega d} \omega^\mu - \frac{g_{\rho d}}{2} \va{\tau} \vdot \va{\rho}^\mu} - \pqty{m_d - g_{\sigma d} \sigma}} \Psi_d \nonumber                                                                                                     \\
                  & + \frac{1}{2} \pqty{\partial_{\mu}\sigma\partial^{\mu}\sigma -m_\sigma^2 \sigma^2} - \frac{1}{4}\Omega_{\mu\nu}\Omega^{\mu\nu} +\frac{1}{2} m_\omega^2 \omega_{\mu}\omega^{\mu} - \frac{1}{4} \va{R}_{\mu\nu} \vdot \va{R}^{\mu\nu} + \frac{1}{2} m_\rho^2 \va{\rho}_{\mu} \vdot \va{\rho}^\mu \nonumber \\
    \label{eq:lagrangian}
\end{align}
where we have used the Rarita-Schwinger-type Lagrangian density~\cite{lavagno_strangeness_2022} for the $\Delta$-baryons, converting it to the form of a Dirac equation in the mean field approximation~\cite{paoli_2013}. The baryon and lepton masses are represented by $m_i,$ where $\,i \in n,p,l,H,D$, whereas the mesons masses are denoted by $m_\sigma$, $m_\omega$ and $m_\rho$. The $\omega$ and $\rho$ meson field-strength tensors are given by $\Omega_{\mu\nu}=\partial_{\mu}\omega_{\nu}-\partial_{\nu}\omega_{\mu}$ and $\va{R}_{\mu\nu}=\partial_{\mu} \va{\rho}_{\nu}-\partial_{\nu} \va{\rho}_{\mu} - g_\rho (\va{\rho}_\mu \cp \va{\rho}_\nu)$, respectively.

\begin{table}[tbp]
    \centering
    \caption{Parameter values at saturation density ($n_0$) used in the DD-MEX model are listed~\cite{taninah_parametric_2020}. The meson-nucleon couplings for the $\sigma$, $\omega$ and $\rho$ mesons included in the matter composition are given by $g_{\sigma N}$, $g_{\omega N}$ and $g_{\rho N}$, respectively. The $m_\sigma$, $m_\omega$ and $m_\rho$ are the meson masses and are given in units of MeV.}
    \begin{tabular}{|c|cccccc|}
    \hline
    Coupling & $g_{\sigma N}$ & $g_{\omega N}$ & $g_{\rho N}$ & $m_\sigma$ & $m_\omega$ & $m_\rho$ \\
    Model & & & & (MeV) & (MeV) & (MeV) \\
    \hline
    DD-MEX & 10.7067 & 13.3388 & 7.2380 & 547.3327 & 783 & 763 \\
    \hline
    \end{tabular}
    \label{tab:parameters}
\end{table}
\begin{table}[tbp]
    \centering
    \caption{The coefficient values used in the scaling equations (\eqref{eq:g_scaling_1} and \eqref{eq:g_scaling_2}) for the DD-MEX model are listed~\cite{taninah_parametric_2020}.}
    \begin{tabular}{|c|cccc|}
    \hline
    Meson ($i$) & $a_i$     & $b_i$     & $c_i$     & $d_i$ \\
    \hline
    $\sigma$    & 1.3970    & 1.3350    & 2.0671    & 0.4016\\
    $\omega$    & 1.3926    & 1.0191    & 1.6060    & 0.4556\\
    $\rho$      & 0.6202    &           &           &       \\
    \hline
    \end{tabular}
    \label{tab:coefficients}
\end{table}

The coupling constants $g_i$ ($i=\sigma, \omega, \rho$) in the DDRMF model are scaled according to the baryon density ($n_b$) to reproduce the bulk properties of nuclear matter and this scaling is given by~\cite{hofmann_2001},
\begin{equation}
	g _{i}(n_b)=g _{i}(n_0)a_{i}\frac{1+b_{i}{(\eta+d_{i})}^{2}}{1+c_{i}{(\eta+d_{i})}^{2}}\, ,
	\label{eq:g_scaling_1}
\end{equation}
for $i=\sigma,\omega$ and for the $\rho$ meson it is given by
\begin{equation}
	g _{\rho}(n_b)=g _{\rho}(n_0) \exp \qty{-a_{\rho }(\eta-1)}\, ,
	\label{eq:g_scaling_2}
\end{equation}
where $\eta = n_b/n_0$ and $n_0$ is the nuclear saturation density. The parameter values along with the scaling coefficients corresponding to the DD-MEX model are listed in Tables~\ref{tab:parameters} and~\ref{tab:coefficients}.

The values of the hyperon-meson and $\Delta$-meson couplings constants can be obtained by parameterizing them in terms of the nucleon-meson couplings by using the ratio $x_{ib}=g_{ib}/g_{iN}$, with $i = \sigma, \omega, \rho$ and $b = N, H, \Delta$, fixing $x_{iN}$ at $1$. The vector meson-hyperon coupling constants have been shown to be related to the vector meson-nucleon couplings via the SU(6) symmetry group as~\cite{schaffner_multiply_1994, dover_hyperon-nucleus_1984},
\begin{align}
	 & x_{\omega \Lambda} = x_{\omega \Sigma} = \frac{2}{3}\, , \quad x_{\omega \Xi} = \frac{1}{3}\, , \\
	 & x_{\rho \Sigma} = 2\, , \quad x_{\rho \Xi} = 1\, , \quad x_{\rho \Lambda} = 0\,.
	\label{eq:SU6}
\end{align}
The scalar meson-hyperon coupling constants are computed from the hyperon potential depth at saturation density, which is defined as~\cite{glendenning_compact_1996, pais_dynamical_1966, lopes_hypernuclear_2014, schaffner_multiply_1994},
\begin{equation}
	U_H^{(N)} = - g_{\sigma H} \sigma (n_0) + g_{\omega H} \omega (n_0)\, ,
	\label{eq:hyp_pot}
\end{equation}
and the values considered here are $U_\Lambda = - 30$MeV, $U_\Sigma = 30$MeV and $U_\Xi = - 14$MeV.

Owing to the scarcity of conclusive experimental data on how $\Delta$-resonances couple to mesons, we do not impose any constraints when choosing the values for $x_{i\Delta}$ and instead vary them in the following ranges,\footnote{The ranges chosen ensure that the resulting EoSs exhibit a sufficiently broad spectrum of variations which facilitates the examination of the influence of the coupling strengths while minimizing the computational load by generating an optimal number of EoSs.}
\begin{align}
	0.8 \leq x_{\sigma \Delta} \leq 1.2\, , \nonumber \\
	1.0 \leq x_{\omega \Delta} \leq 1.1\, ,           \\
	0.5 \leq x_{\rho \Delta} \leq 1.5\,.
\nonumber
	\label{eq:delta_meson_coup}
\end{align}

In order to satisfy the $ \beta $-equilibrium condition in a NS with baryons and leptons, the chemical potentials of the particles must satisfy the following relations,
\begin{align}
	\mu_{\Sigma^-} = \mu_{\Xi^-}    & = \mu_{\Delta^-} = \mu_n + \mu_e\, ,       \\
	\mu_{\mu}                       & = \mu_e\, ,                                \\
	\mu_\Lambda = \mu_{\Sigma^0}    & = \mu_{\Xi^0} = \mu_{\Delta^0} = \mu_n\, , \\
	\mu_{\Sigma^+} = \mu_{\Delta^+} & = \mu_p = \mu_n - \mu_e\, ,                \\
	\mu_{\Delta^{++}}               & = 2\mu_p - \mu_n\,.
\end{align}
These chemical potentials are given by,
\begin{align}
	\mu_b & = \sqrt{{k_F^b}^2 + {m_b^\ast}^2} + g_{\omega b} \omega + g_{\rho b} \tau_{3 b} \rho + \Sigma^r\, , \\
	\mu_d & = \sqrt{{k_F^d}^2 + {m_d^\ast}^2} + g_{\rho d} \tau_{3 b} \rho + \Sigma^r\, ,                       \\
	\mu_l & = \sqrt{{k_F^l}^2 + m_l^2}\, ,
\end{align}
where $ k_F $ is the Fermi momentum of the particle, $\tau_{3b}$ is the isospin projection of the baryon and $ \Sigma^r $ is a rearrangement term arising due to the density-dependent couplings given by,
\begin{equation}
	\Sigma^r = \sum_b \bqty{\frac{\partial g_{\omega b}}{\partial n_b} \omega n_b + \frac{\partial g_{\rho b}}{\partial n_b} \rho \tau_{3 b} n_b - \frac{\partial g_{\sigma b}}{\partial n_b} \sigma n_b^s + b\leftrightarrow d}\,.
\end{equation}
Here $ m^\ast_b $ and $ m^\ast_d $ are the Dirac effective masses given by,
\begin{equation}
	m^\ast_b = m_b - g_{\sigma b} \sigma\, , \quad m^\ast_d = m_d - g_{\sigma d} \sigma\, ,
	\label{eq:effect_mass}
\end{equation}
and $ n_i^s\, \pqty{ i \in b,\, d } $ is the scalar density given by,~\cite{lopes_hypernuclear_2014}
\begin{equation}
	n_i^s = \gamma_i \int\limits_{0}^{k_F^i} \frac{m_i^\ast}{\sqrt{k^2 + {m_i^\ast}^2}} \frac{k^2}{2\pi^2} \dd{k} \,,
	\label{eq:scalar_density}
\end{equation}
where $\gamma_i$ is the spin degeneracy parameter. Alongside the chemical equilibrium condition, the NS matter also needs to satisfy charge neutrality condition which is imposed by the equation,
\begin{equation}
	n_p + n_{\Sigma^+} + 2n_{\Delta^{++}} + n_{\Delta^+} = n_{\Sigma^-} + n_{\Xi^-} + n_{\Delta^-} + n_e + n_\mu\,.
	\label{eq:charge_neutrality}
\end{equation}
The equations of motion of the mesons are obtained using the relativistic mean-field approximation,
\begin{align}
	m_\sigma^2 \sigma & = \sum_b g_{\sigma b} n_b^s + \sum_d g_{\sigma d} n_d^s\, ,            \\
	m_\omega^2 \omega & = \sum_b g_{\omega b} n_b + \sum_d g_{\omega d} n_d\, ,                \\
	m_\rho^2 \rho     & = \sum_b g_{\rho b} n_b \tau_{3b} + \sum_d g_{\rho d} n_d \tau_{3d}\,.
	\label{eq:field_equations}
\end{align}
The energy density of the system can be written as,
\begin{align}
	\varepsilon = & \sum_{i \in b, \Delta} \frac{\gamma_i}{2\pi^2} \int\limits_{0}^{k_F^i} k^2 \sqrt{{m^\ast_i}^2 + k^2} \dd{k} \nonumber \\
	              & + \sum_l \frac{1}{\pi^2} \int\limits_{0}^{k_F^l} k^2 \sqrt{m_l^2 + k^2} \dd{k} \nonumber                                    \\
	              & + \frac{1}{2} \pqty{m_\sigma^2 \sigma^2 + m_\omega^2 \omega^2 + m_\rho^2 \rho^2}\, ,
	\label{eq:energy}
\end{align}
while the pressure is given by,
\begin{align}
	P = & \sum_{i \in b,\Delta} \frac{\gamma_i}{3 \pqty{2\pi^2}} \int\limits_{0}^{k_F^i} \frac{k^4}{\sqrt{k^2 + {m^\ast_i}^2}} \dd{k} \nonumber \\
	    & + \sum_l \frac{1}{3\pi^2} \int\limits_{0}^{k_F^l} \frac{k^4}{\sqrt{k^2 + m_l^2}} \dd{k} + n_b \Sigma^r \nonumber \\
	    & + \frac{1}{2} \pqty{-m_\sigma^2 \sigma^2 + m_\omega^2 \omega^2 + m_\rho^2 \rho^2}\,.
	\label{eq:pressure}
\end{align}

\section{Results and Discussion}\label{results}
\begin{figure*}
	\centering
	\includegraphics[width=\linewidth, keepaspectratio]{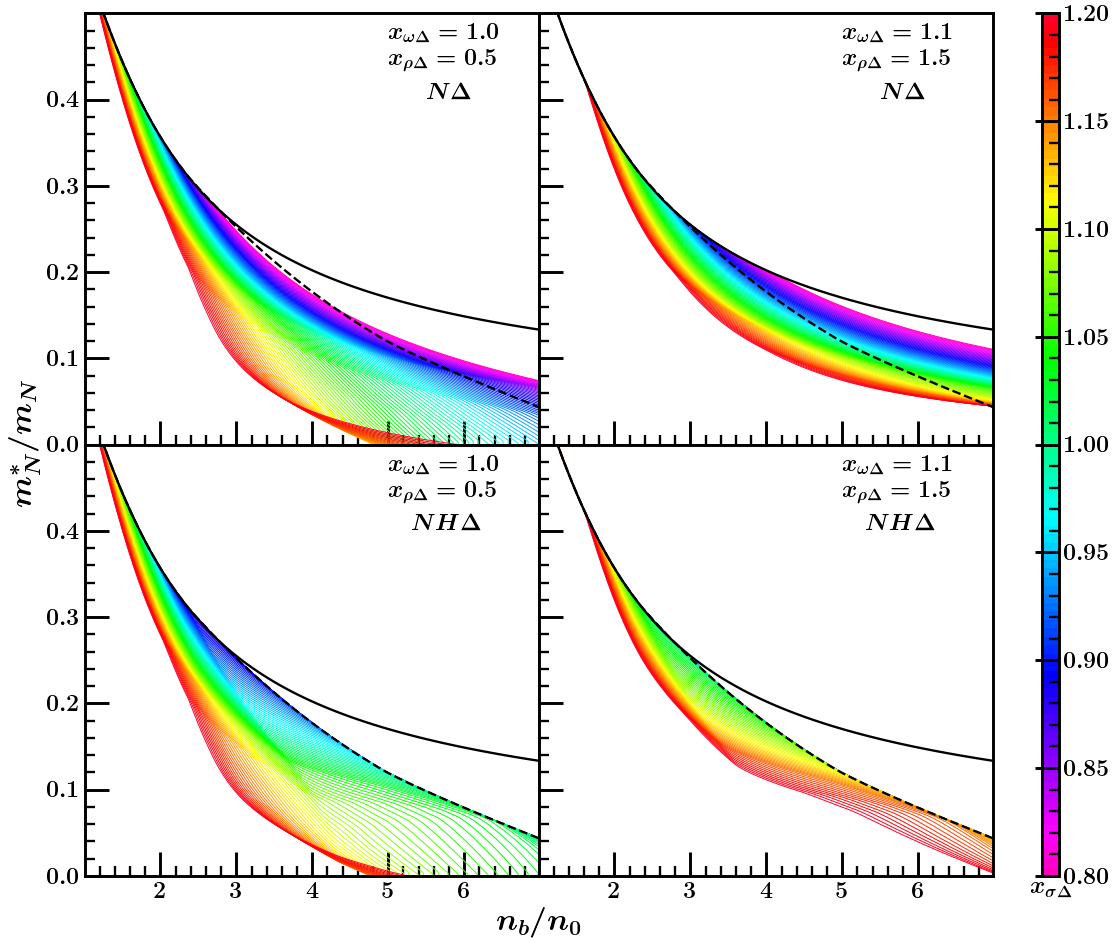}
	\caption{Normalized nucleon effective mass as a function of density for hyperon-free $\Delta$-admixed NS matter (upper half) and $\Delta$-admixed hypernuclear matter (lower half). The sub-figures on the left correspond to the combination of $x_{\omega\Delta}=1.0$ and $x_{\rho\Delta} = 0.5$, while the ones on the right correspond to $x_{\omega\Delta}=1.1$ and $x_{\rho\Delta} = 1.5$, respectively. The $\sigma-\Delta$ coupling strength is varied in the range $x_{\sigma\Delta}\in [0.8, 1.2]$ in all the sub-figures (corresponding colour given in the colour-bar on the right). The solid black line in the sub-figures is the $m_N^*/m_N$ of NS matter composed of only nucleons and leptons, whereas the dashed black line is for NS matter containing nucleons, leptons and hyperons.}\label{fig:effective_mass}
\end{figure*}
\begin{figure*}
    \centering
    \includegraphics[width=\linewidth, keepaspectratio]{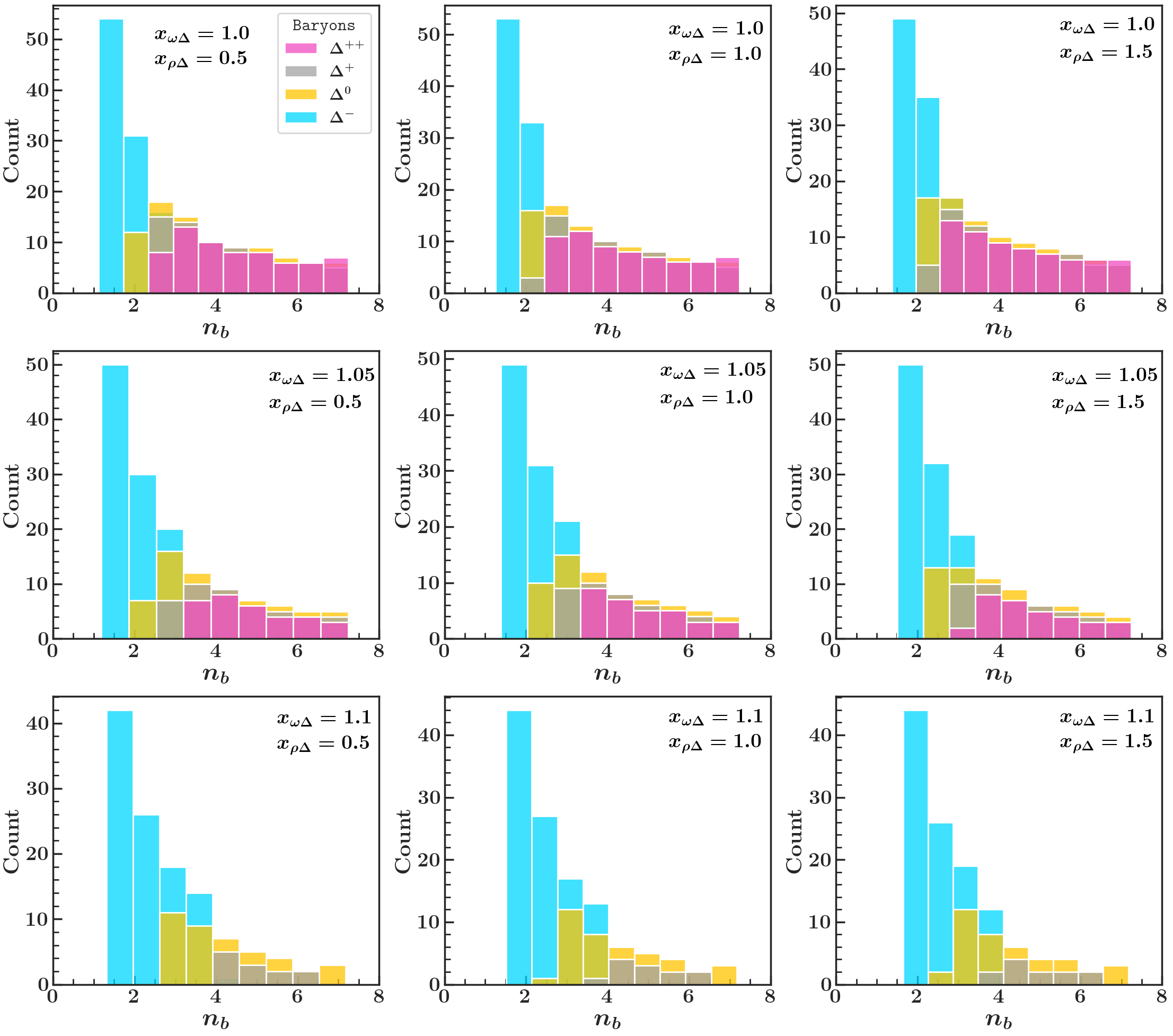}
    \caption{Histograms depicting the distribution of the threshold densities of the $\Delta$-baryons in the NS matter. The sub-figures correspond to the different possible combinations of $x_{\omega\Delta}$ and $x_{\rho\Delta}$. The different baryons are represented using the colors in the legend provided in the first sub-figure.}
    \label{fig:population_ND}
\end{figure*}

We begin by exploring the characteristics of heavy baryons within NSs. The DD-MEX model, which was introduced in the previous section, provides a microscopic approach towards understanding the composition and possibility of occurrence of various heavy baryons in charge-neutral, $\beta$-stable NS matter at zero temperature. In this work, we focus on understanding the behaviour of NS oscillations when $\Delta$-baryons are present in the NS matter composition. Additionally, considering that hyperons can also nucleate in the NS core, as we saw in section~\ref{sec:Introduction}, it becomes essential to take their presence into account. This is done by taking two different NS matter compositions in our study, one being a hyperon-free NS matter containing nucleons, leptons and $\Delta$-baryons (N$\Delta$) only, and the other being $\Delta$-admixed hypernuclear matter (NH$\Delta$) containing nucleons, leptons, hyperons and $\Delta$-baryons.

\begin{figure*}
	\centering
	\includegraphics[width=\linewidth, keepaspectratio]{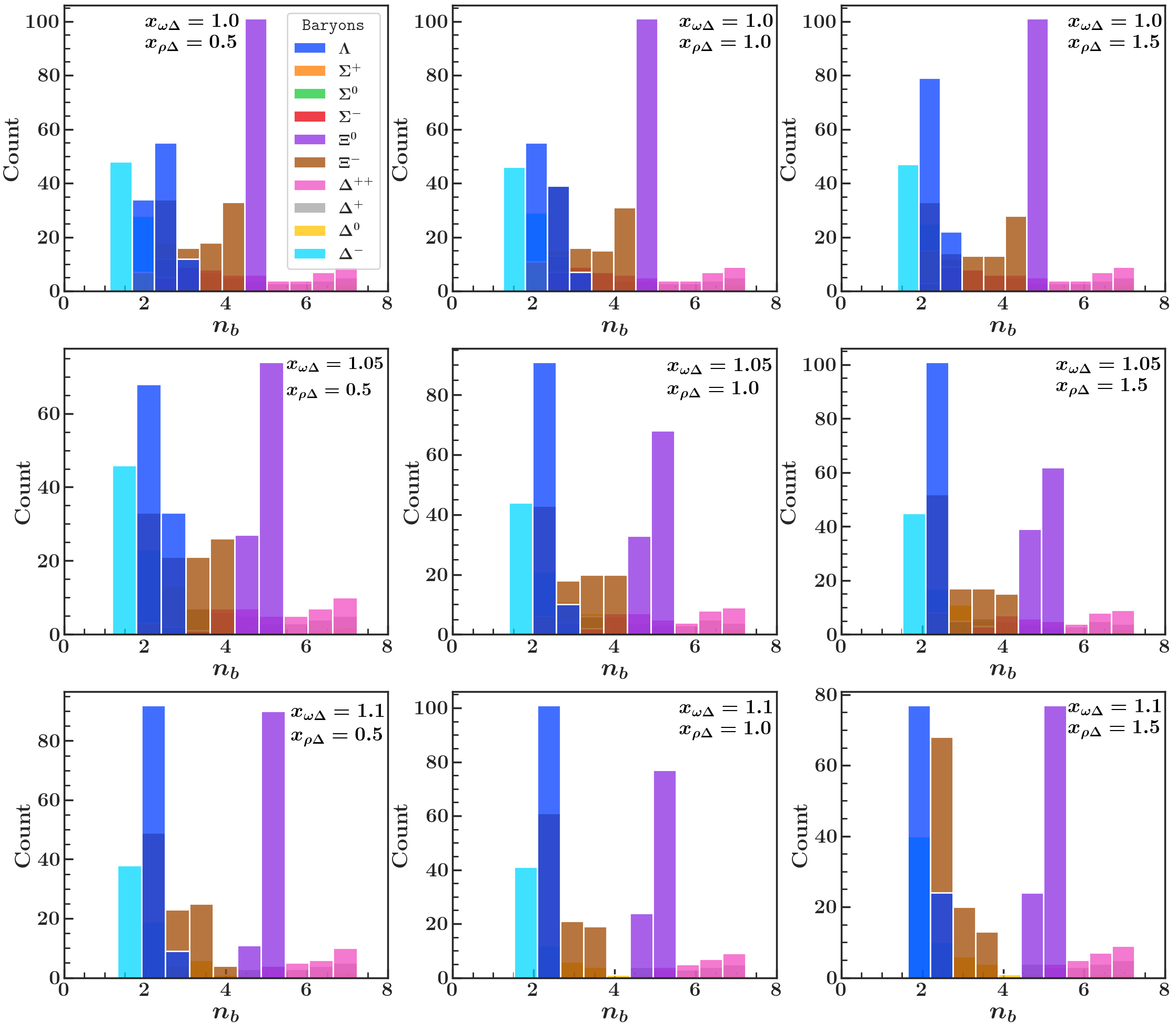}
	\caption{Similar to figure~\ref{fig:population_ND} but for $\Delta$-admixed hypernuclear matter.}\label{fig:population_NYD}
\end{figure*}

The behaviour of the nucleon effective mass as a function of the baryon density is a topic of significant interest when studying NS properties~\cite{zhang_massive_2018, jaiswal_constraining_2021}, such as the mass-radius relations and $f$-mode frequencies~\cite{marquez_delta_2022}. The nucleon effective mass ($m_N^\ast$) decreases with increasing baryon density $n_b$, and we see that in the absence of any other baryonic species, the rate of decrease declines gradually with $n_b$ such that $m_N^\ast/m_N$ does not fall below $0.1$ even in the high density regime. Addition of other baryonic species, such as hyperons or $\Delta$-resonances, causes the nucleon effective mass to decrease at a much faster rate due to the additional negatively contributing term from the scalar density dependence of the $\sigma$ field in eq.~\eqref{eq:effect_mass}. In the figure~\ref{fig:effective_mass}, we plot the normalized nucleon effective mass as a function of density to illustrate the effect of different baryons being present in the matter composition. We find that, keeping in agreement with the results obtained by Marquez et. al.\cite{marquez_delta_2022}, the value of $m_N^\ast$ decreases to zero (at baryon densities above $4.5 n_0$) for certain combinations of $x_{b\Delta}$. This leads to the possibility that the nucleon effective mass could become zero at some density before the NS maximum mass configuration is reached. This can be solved by considering a phase transition to some exotic matter composition occurring at some density before $m_N^\ast$ reaches zero, which is beyond the scope of the current work. Contrarily, we find that the rate of decrease becomes less drastic for higher values of meson-$\Delta$ coupling constants, thereby leading to certain cases where $m_N^\ast$ does not approach zero for any of the values of $x_{\sigma\Delta}$ considered here. To gain a deeper insight into the influence of the various particle species on the properties of NSs, we examine the threshold nucleation density of the different baryons considered. Figures~\ref{fig:population_ND} and~\ref{fig:population_NYD} present the plots for the threshold density at which these particles first appear in the system. The histograms show the effect that varying $x_{\sigma\Delta} \in [0.8, 1.2]$ has on the threshold density of each baryonic species.

In $\Delta$-admixed NS matter~(figure~\ref{fig:population_ND}), we observe that after nucleons and leptons, the first particle to appear is the negatively charged $\Delta^-$ baryon, which emerges (on average) near the $2n_0$ mark. The charge neutrality condition imposed on the NS matter suppresses the presence of positively charged $\Delta^+$ baryons, leading to the absence of $\Delta^{++}$ baryons in combinations where $x_{\omega\Delta} = 1.1$. Furthermore, we find that even $\Delta^0$ or $\Delta^+$ do not nucleate for all $x_{\sigma\Delta}$ values and require stronger meson-baryon couplings to do so, which pushes the nucleation threshold to higher densities causing the baryons to be located well inside the NS core. Moving on to $\Delta$-admixed hypernuclear matter (figure~\ref{fig:population_NYD}), we observe that the only hyperons present in the system are $\Lambda$ and $\Xi^{0,-}$. Similar to the N$\Delta$ matter case, higher values of $x_{\omega\Delta}$ have a comparable impact on the $\Delta$-baryons, causing them to appear at higher average densities. These results highlight that enforcing charge neutrality significantly favors the emergence of negatively charged baryons, with the spin-$3/2$ $\Delta^-$ being the most favored. The preference for $\Delta^-$ over the lighter, neutrally charged $\Lambda$ can be attributed to the more attractive potential of $\Delta^-$ which can overcome the mass difference when replacing a neutron-electron pair.
Moreover, analysing figure~\ref{fig:population_NYD} shows us that $x_{\omega\Delta}$ impacts the hyperon threshold density by supporting the nucleation of hyperons at lower densities while at the same time suppressing the $\Delta$-baryons from nucleating when $x_{\omega\Delta}$ is increased.

\begin{figure*}
	\centering
	\includegraphics[width=\linewidth, keepaspectratio]{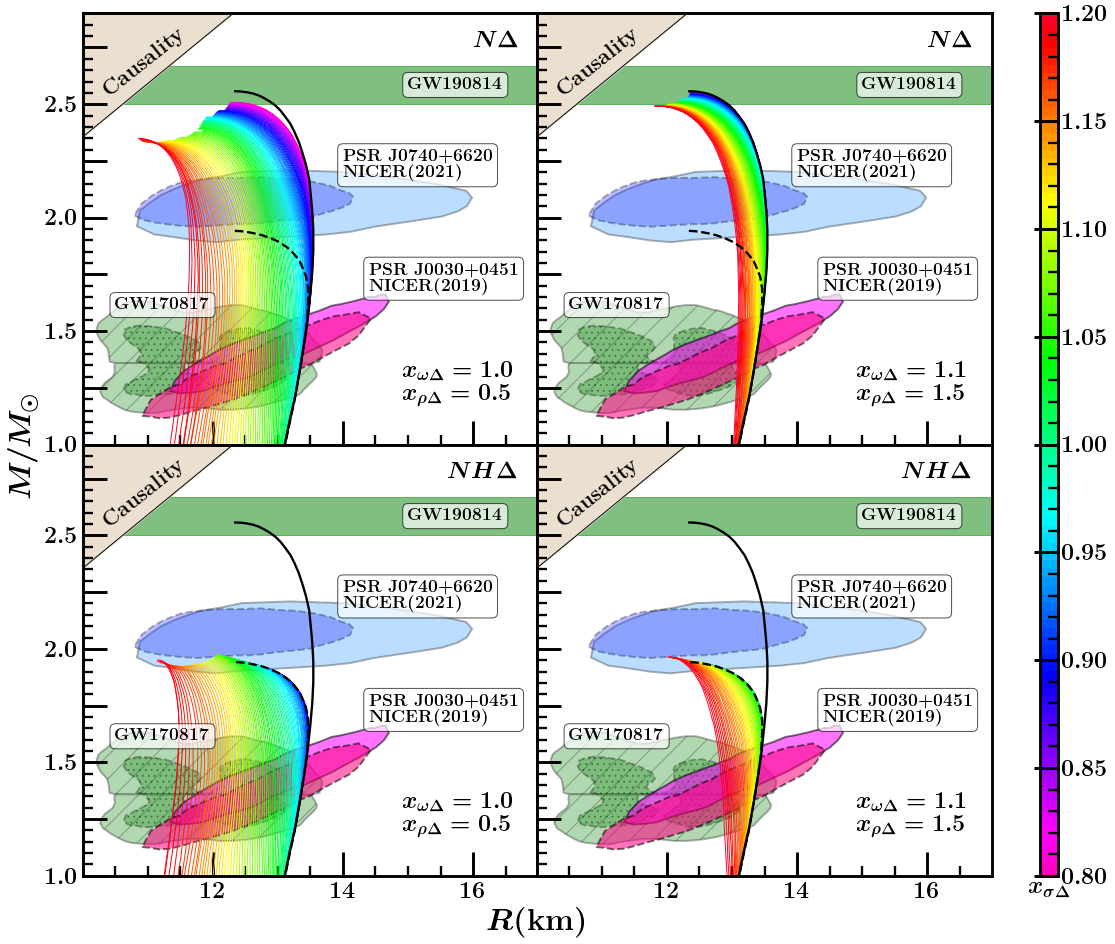}
	\caption{Mass-radius curves showing the effect of varying $x_{\sigma\Delta}$ with different combinations of $x_{\omega\Delta}$ and $x_{\rho\Delta}$ for hyperon-free $\Delta$-admixed NS matter (upper half) and $\Delta$-admixed hypernuclear matter (lower half). The solid and dashed black lines represent compositions of NS matter corresponding to nucleons and leptons, and nucleons, leptons and hyperons respectively. The value of $x_{\sigma\Delta}$ taken for each curve is represented by the corresponding colour given in the color-bar on the right. The horizontal green band at the top is the mass constraint obtained from the gravitational wave event GW190814~\cite{abbott_gw190814_2020}, while the green region with shading located at the bottom left is the constraint obtained from GW170817~\cite{abbott_gw170817_2017}. The constraints on mass and radius obtained from pulsars is given by the pink region for PSR J0030+0451 from the 2019 NICER data~\cite{riley_nicer_2019, miller_psr_2019}, and by the blue region for PSR J0740+6620 from the 2021 NICER data~\cite{riley_nicer_2021, fonseca_refined_2021}.}\label{fig:mass_radius}
\end{figure*}

In this study, we applied the Tolman-Oppenheimer-Volkoff (TOV) equations of relativistic hydrostatic equilibrium \cite{tolman_static_1939, oppenheimer_massive_1939} to derive families of stars based on the equations of state (EoS) generated for different combinations of $x_{\sigma\Delta}$, $x_{\omega\Delta}$, and $x_{\rho\Delta}$. The EoSs used are unified, meaning that the core EoSs computed by us are supplemented by the SLY4 EoS~\cite{douchin_unified_2001} for the low density crust region, and the crust-core matching is done by requiring both pressure and density to be continuous at the point of matching. The corresponding families of stars are illustrated in figure~\ref{fig:mass_radius} for hyperon-free NS matter containing $\Delta$-baryons in the upper half and for $\Delta$-admixed hypernuclear matter in the lower half.
The color bar accompanying the figures indicates the varied $x_{\sigma\Delta}$ values within the range $[0.8,1.2]$. To compare the effects of including additional baryons into the system, we also plot the mass-radius curves for NS matter containing only nucleons and leptons (solid black line) and hypernuclear matter containing nucleons, leptons and hyperons (dashed black line). All the mass-radius curves in the figure are plotted up to the maximum mass configuration obtained from their corresponding EoSs. We have also included constraints on the mass and radius of neutron stars from multiple observational sources in the figure. The green horizontal band corresponds to constraints derived from the gravitational wave event GW190814~\cite{abbott_gw190814_2020}, while the green shaded region located towards the bottom left corresponds to the gravitational wave event GW170817~\cite{abbott_gw170817_2017}. The two pink regions represent constraints obtained from 2019 NICER data of the pulsar PSR J0030+0451~\cite{riley_nicer_2019, miller_psr_2019}, while the blue regions depict constraints from 2021 NICER data of the pulsar PSR J0740+6620~\cite{riley_nicer_2021, fonseca_refined_2021}. Despite the considerable uncertainties in the measurements, our models demonstrate agreement with the observational constraints for various matter composition scenarios, whether with nucleons and $\Delta$'s or with the inclusion of hyperons.

\begin{figure*}
	\centering
	\includegraphics[width=\textwidth, keepaspectratio]{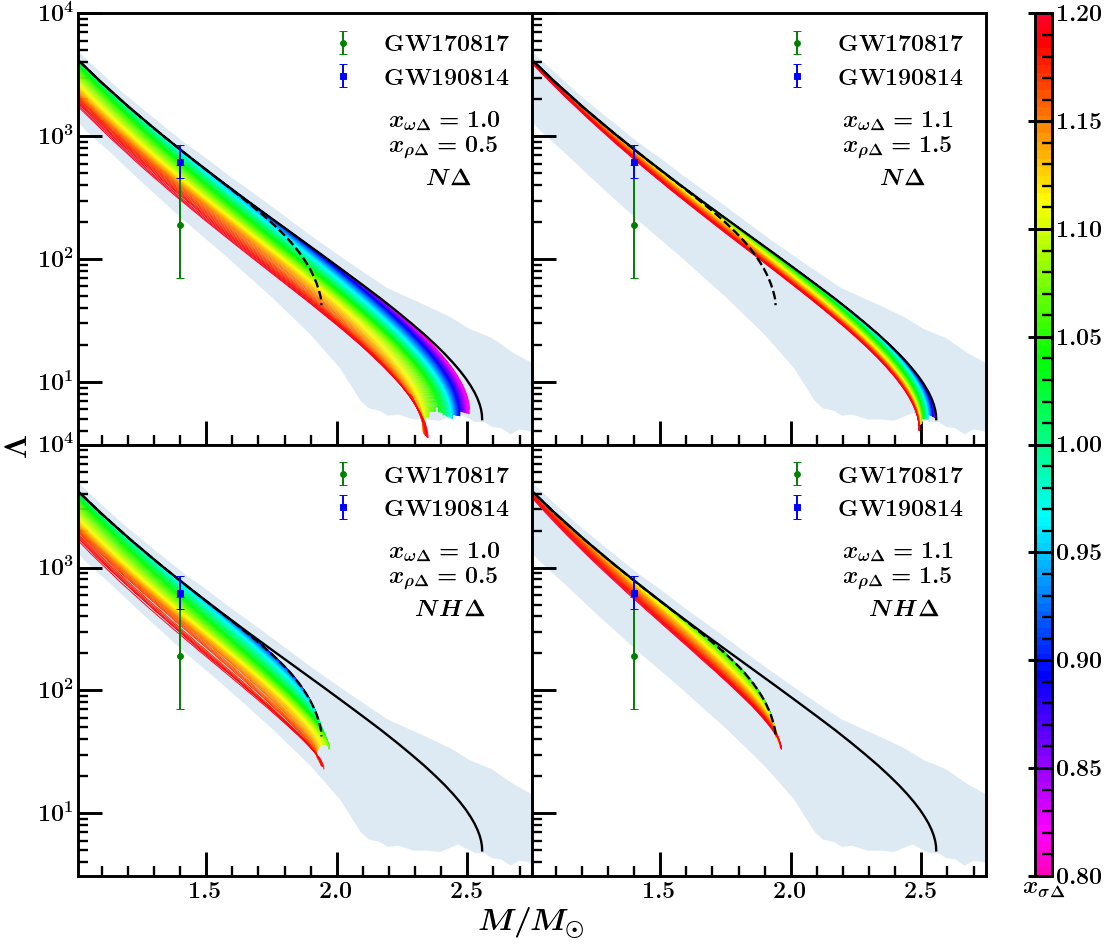}
	\caption{Dimensionless tidal deformability ($\Lambda$) against NS mass for $\Delta$-admixed NS matter (upper half) and $\Delta$-admixed hypernuclear matter (lower half), showing the effect of varying $x_{\sigma\Delta}$ with different combinations of $x_{\omega\Delta}$ and $x_{\rho\Delta}$. To represent the different $x_{\sigma\Delta}$ values, we use the corresponding color given in the adjoining color-bar. A solid black line is used to represent NS matter containing nucleons and leptons only, whereas the dashed black line is for NS matter containing nucleons, hyperons and leptons only. Observational constraints are represented by the green error-bar and grey shaded patch for GW170817~\cite{abbott_gw170817_2017}, and the blue error-bar for GW190814~\cite{abbott_gw190814_2020}.}\label{fig:tidal_exp}
\end{figure*}

From the figure, we can infer that the EoS of NSs is affected by the various couplings between the mesons and baryons in the system. In particular, we find that the $\Delta$-resonances can play a vital role with the coupling constants $x_{\sigma\Delta}$, $x_{\omega\Delta}$, and $x_{\rho\Delta}$ being the most relevant. The impact of these couplings on the stellar radius is shown in the figure where we observe a decrease in the star's radius upon increasing the $x_{\sigma\Delta}$ strength, as the attraction increases and the EoS softens at intermediate densities. Similarly, decreasing $x_{\rho\Delta}$ results in smaller radii since this reduces the repulsion associated with proton-neutron asymmetry. Importantly, we note that the presence of hyperons and $\Delta$'s together can increase the maximum mass limit beyond that of hyperonic matter if $x_{\omega\Delta}\geq 1$ which can be attributed to the vector meson dominating at high densities and its coupling with the $\Delta$-baryon is stronger compared to its nucleon or hyperon couplings. The relationship between these couplings and the maximum mass limit is complex and requires further discussion to be fully understood.

\begin{figure*}
	\centering
	\includegraphics[width=\linewidth, keepaspectratio]{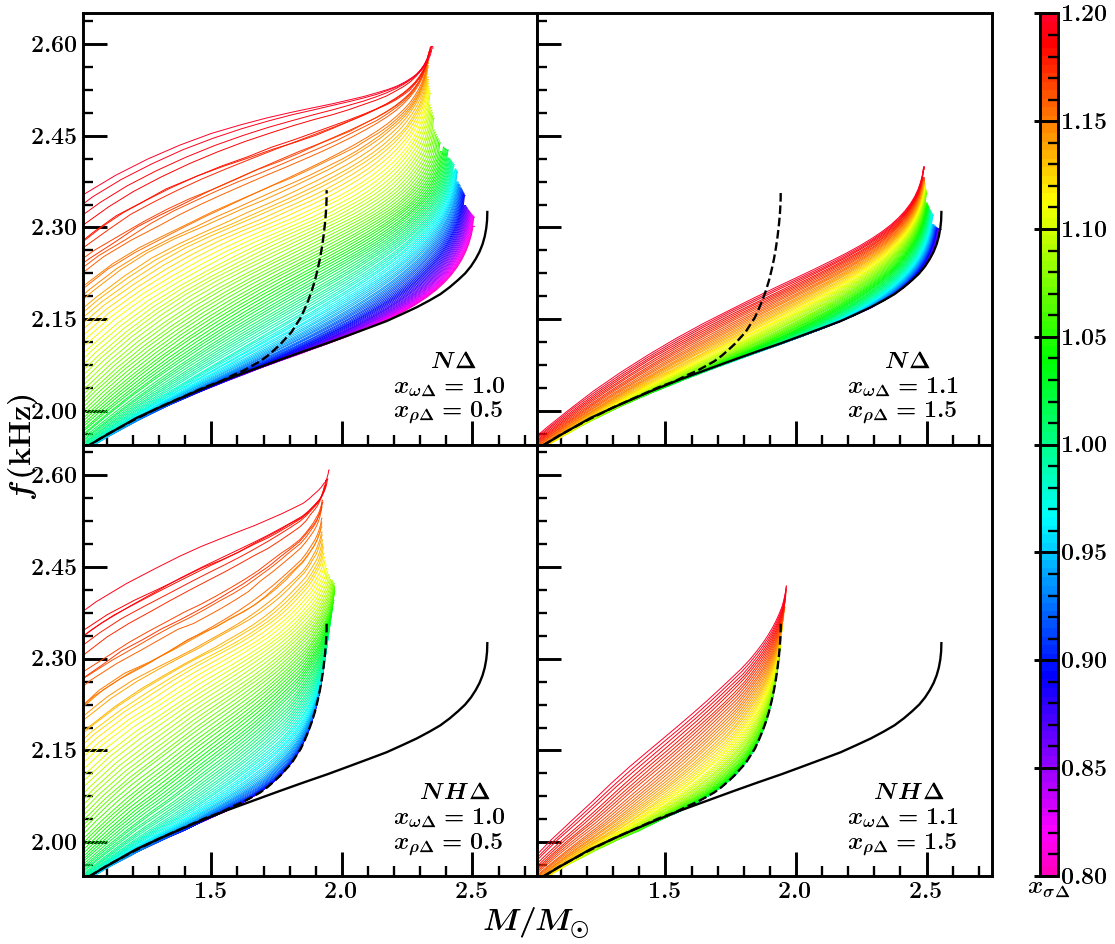}
	\caption{$f$-mode oscillation frequency against NS mass for $\Delta$-admixed NS matter (upper half) and $\Delta$-admixed hypernuclear matter (lower half), showing the effect of varying $x_{\sigma\Delta}$ with different combinations of $x_{\omega\Delta}$ and $x_{\rho\Delta}$. To represent the different $x_{\sigma\Delta}$ values, we use the corresponding color given in the adjoining color-bar. A solid black line is used to represent NS matter containing nucleons and leptons only, whereas the dashed black line is for NS matter containing nucleons, hyperons and leptons only.}\label{fig:freq_exp}
\end{figure*}

When present in a binary system, NSs experience tidal effects caused by the companion's gravitational field. These effects can be quantified by means of the dimensionless tidal deformability ($ \Lambda $), defined as $\Lambda = \frac{2}{3} k_2 C^{-5}$, where $ k_2 $ is the tidal love number and C is the compactness \cite{hinderer_tidal_2008, hinderer_tidal_2010, flanagan_constraining_2008}. We investigate $\Lambda$ in two scenarios: (1) for $\Delta$-admixed NS matter (upper half of figure~\ref{fig:tidal_exp}) and (2) for $\Delta$-admixed hypernuclear matter (lower half of figure~\ref{fig:tidal_exp}). In both cases, we explore various combinations of $x_{\omega\Delta}$ and $x_{\rho\Delta}$, while varying $x_{\sigma\Delta}$. To distinguish between nuclear and hypernuclear matter compositions, we use black solid and dashed lines, respectively, in our plots. Additionally, we include observational constraints on tidal deformability at the canonical mass ($1.4M_\odot$) from the gravitational wave events GW170817~\cite{abbott_gw170817_2017} and GW190814~\cite{abbott_gw190814_2020}. Our findings indicate that, at constant NS mass, increasing the coupling between the $\sigma$ meson and the $\Delta$-baryons leads to a decrease in $\Lambda$ compared to the scenario with nucleon-only NS matter. This reduction can be attributed to an increase in the attractive interactions thereby causing the EoS to soften. However, we observe that this decrease in $\Lambda$ can be mitigated by enhancing the $\omega-\Delta$ and $\rho-\Delta$ coupling strengths, which promotes repulsive interactions among the $\Delta$-baryons. In the case of $\Delta$-admixed hypernuclear matter, we find that the band of $\Lambda$ at a given mass gets shifted downwards due to the attractive interactions arising from the presence of hyperons. Remarkably, for values of $x_{\sigma\Delta}$ greater than 1, the NS exhibits a significantly lower $\Lambda$ value than in the scenario with only nucleons and hyperons ($NH$ only case). Additionally, we observe that increasing $x_{\omega\Delta}$ above 1 has a noticeable effect on the maximum mass in this context.

NS oscillations arising due to perturbations (either external or internal), cause emission of gravitational waves. These waves are emitted in different frequency modes with the fundamental mode being denoted by $ f $. Cowling approximation~\cite{cowling_non-radial_1941,sotani_signatures_2011, vasquez_flores_2014, ranea-sandoval_2018, das_impacts_2021} is one of the most popular methods of solving the equations for non-radial oscillations. Using figure~\ref{fig:freq_exp}, we study the influence of meson-$\Delta$ baryon interactions on the non-radial $f$-mode oscillation frequency for NSs composed of $\Delta$-admixed NS matter (upper half of figure \ref{fig:freq_exp}) and $\Delta$-admixed hypernuclear matter (lower half of figure  \ref{fig:freq_exp}). Consistent with the previous figures, the solid and dashed black lines represent $N$ and $NH$ matter compositions, respectively. We observe from figure~\ref{fig:mass_radius} that as we progressively increase the coupling strength between the $\sigma$ meson and the $\Delta$-baryons, the resulting star is able to attain a smaller radius and lower mass. Consequently, the $f-$mode frequency is is seen to increase, as is evident in figure~\ref{fig:freq_exp}. Additionally, we observe that lower values of $x_{\omega\Delta}$ and $x_{\rho\Delta}$ lead to a wider variation in the $f-$mode frequency at constant mass, particularly in the low mass region. This variation is attributed to the presence of a greater number of $\Delta$-baryons in the NS core, resulting from the larger attractive interaction and smaller repulsive interaction. Conversely, higher values of $x_{\omega\Delta}$ and $x_{\rho\Delta}$ significantly compress the range of $f$-mode frequencies in the low mass region for a given mass, owing to the dominance of repulsive interactions. These observations are consistent with the effects of meson interactions on the NS radius, as shown in figure~\ref{fig:mass_radius}. Furthermore, we find that similar to the case of dimensionless tidal deformability, presence of hyperons also impacts the variation of $f-$mode frequency at a given mass - increasing the $f-$mode frequency significantly, especially for $x_{\sigma\Delta} \geq 1$.

\section{Conclusion}\label{conclusion}
In this study, we attempted to understand the impacts of heavy baryons on NS properties, while keeping them constrained using available observational data. To achieve this, we utilized the DD-MEX model within the Density-Dependent Relativistic Mean Field (DDRMF) framework, enabling a systematic exploration of how $\Delta$-admixed hypernuclear and hyperon-free NS matter influences NS oscillations. This approach allowed us to reveal insights into particle compositions, their emergence processes, and their profound influence on key NS properties. We investigated the effects of heavy baryons on the non-radial $f$-mode oscillations of NSs, discovering a direct correlation between the frequency of the oscillation mode and the $\sigma$-$\Delta$ coupling strength. This correlation manifested in the coupling's impact on stellar mass and radius. The repulsive $x_{\omega\Delta}$ and $x_{\rho\Delta}$ couplings were also identified as contributors to frequency variation, particularly for low-mass NSs, aligning with our observations of meson interactions effects on NS radii.

The variation in the fundamental mode oscillation frequency of NSs was attributed to the presence of $\Delta$-baryons in the core. Our analysis of nucleation threshold densities revealed that a larger $x_{\sigma\Delta}$ value, coupled with smaller $x_{\omega\Delta}$ and $x_{\rho\Delta}$ values, favored increased $\Delta$-baryons nucleation in the stellar core, and vice versa. This perspective offers a novel understanding of how these resonances impact NS properties and unveils some underlying dynamics. Calculations of the Dirac effective mass of nucleons ($m_N^\ast$) demonstrated a significant influence of various baryonic species, especially the four $\Delta$-baryons. Consistent with existing literature, the introduction of these baryons led to the nucleon effective mass reaching zero as density increased, suggesting intriguing possibilities regarding phase transitions in stars.

Notably, some configurations deviated from the trend of approaching zero effective nucleon mass even at extremely high baryon densities. Imposing the charge neutrality condition resulted in negatively charged baryons (particularly $\Delta^-$) being more likely to nucleate than neutral or positively charged heavy baryons. Specifically, the spin-3/2 particle $\Delta^-$ exhibited excess attractive potential, favoring its nucleation over the lighter and neutral $\Lambda$-hyperon when replacing a neutron-electron pair. The effect of meson-baryon couplings, especially those of $\Delta$-baryons, on the equation of state and, by extension, the radius and maximum mass configuration of NSs emerged as a key insight. The intricate interplay of these factors led to considerable variation between NS models, with $x_{\sigma\Delta}$ having the most significant impact on the equation of state's softening, particularly in the intermediate density regime. Through the incorporation of observational constraints, we demonstrated a remarkable degree of agreement between the models and currently available data, validating our findings regarding the different matter compositions of $N\Delta$ and $NH\Delta$.

Furthermore, our exploration extended to the dimensionless tidal deformability ($\Lambda$), a key parameter for understanding the interior composition of NSs. Our results indicated that the value of $\Lambda$ is directly influenced by the attractive and repulsive interactions within stellar matter, dependent on the strength of couplings between mesons and $\Delta$-resonances. The effects of these interactions were most pronounced in the low-mass region, with $\Lambda$ for the canonical star decreasing by nearly $\sim 70\%$ in some configurations.

\section*{Acknowledgements}\label{acknowledgements}
P.J.K. thanks Khokan Singha and Sailesh Ranjan Mohanty for their help with the computations. Authors thank Prof. Constan{\c{c}}a Provid{\^e}ncia for her insightful discussion that enhanced the depth and quality of our work. B.K. acknowledges partial support from the Department of Science and Technology, Government of India, with grant no. CRG/2021/000101.

\bibliographystyle{JHEP}
\bibliography{bibliography_file.bib}

\providecommand{\href}[2]{#2}\begingroup\raggedright\begin{thebibliography}{100}

\bibitem{tsang_resonant_2012}
D.~Tsang, J.S.~Read, T.~Hinderer, A.L.~Piro and R.~Bondarescu, \emph{Resonant
  {Shattering} of {Neutron} {Star} {Crusts}},
  \href{https://doi.org/10.1103/PhysRevLett.108.011102}{\emph{Phys. Rev. Lett.}
  {\bfseries 108} (2012) 011102}.

\bibitem{franco_quaking_2000}
L.M.~Franco, B.~Link and R.I.~Epstein, \emph{Quaking {Neutron} {Stars}},
  \href{https://doi.org/10.1086/317121}{\emph{ApJ} {\bfseries 543} (2000) 987}.

\bibitem{hinderer_effects_2016}
T.~Hinderer, A.~Taracchini, F.~Foucart, A.~Buonanno, J.~Steinhoff, M.~Duez
  et~al., \emph{Effects of {Neutron}-{Star} {Dynamic} {Tides} on
  {Gravitational} {Waveforms} within the {Effective}-{One}-{Body} {Approach}},
  \href{https://doi.org/10.1103/PhysRevLett.116.181101}{\emph{Phys. Rev. Lett.}
  {\bfseries 116} (2016) 181101}.

\bibitem{chirenti_gravitational_2017}
C.~Chirenti, R.~Gold and M.C.~Miller, \emph{Gravitational {Waves} from
  {F}-modes {Excited} by the {Inspiral} of {Highly} {Eccentric} {Neutron}
  {Star} {Binaries}},
  \href{https://doi.org/10.3847/1538-4357/aa5ebb}{\emph{ApJ} {\bfseries 837}
  (2017) 67}.

\bibitem{thorne_non-radial_1967}
K.S.~Thorne and A.~Campolattaro, \emph{Non-{Radial} {Pulsation} of
  {General}-{Relativistic} {Stellar} {Models}. {I}. {Analytic} {Analysis} for
  {L} $\geq$ 2}, \href{https://doi.org/10.1086/149288}{\emph{ApJ} {\bfseries
  149} (1967) 591}.

\bibitem{thorne_nonradial_1969}
K.S.~Thorne, \emph{Nonradial {Pulsation} of {General}-{Relativistic} {Stellar}
  {Models}.{IV}. {The} {Weakfield} {Limit}},
  \href{https://doi.org/10.1086/150259}{\emph{The Astrophysical Journal}
  {\bfseries 158} (1969) 997}.

\bibitem{thorne_nonradial_1969-1}
K.S.~Thorne, \emph{Nonradial {Pulsation} of {General}-{Relativistic} {Stellar}
  {Models}. {III}. {Analytic} and {Numerical} {Results} for {Neutron} {Stars}},
  \href{https://doi.org/10.1086/150168}{\emph{ApJ} {\bfseries 158} (1969) 1}.

\bibitem{price_non-radial_1969}
R.~Price and K.S.~Thorne, \emph{Non-{Radial} {Pulsation} of
  {General}-{Relativistic} {Stellar} {Models}. {II}. {Properties} of the
  {Gravitational} {Waves}}, \href{https://doi.org/10.1086/149857}{\emph{The
  Astrophysical Journal} {\bfseries 155} (1969) 163}.

\bibitem{zhao_universal_2022}
T.~Zhao and J.M.~Lattimer, \emph{Universal relations for neutron star $f$-mode
  and $g$-mode oscillations},
  \href{https://doi.org/10.1103/PhysRevD.106.123002}{\emph{Phys. Rev. D}
  {\bfseries 106} (2022) 123002}.

\bibitem{zhao_quasinormal_2022}
T.~Zhao, C.~Constantinou, P.~Jaikumar and M.~Prakash, \emph{Quasinormal $g$
  modes of neutron stars with quarks},
  \href{https://doi.org/10.1103/PhysRevD.105.103025}{\emph{Phys. Rev. D}
  {\bfseries 105} (2022) 103025}.

\bibitem{sotani_universal_2021}
H.~Sotani and B.~Kumar, \emph{Universal relations between the quasinormal modes
  of neutron star and tidal deformability},
  \href{https://doi.org/10.1103/PhysRevD.104.123002}{\emph{Phys. Rev. D}
  {\bfseries 104} (2021) 123002}.

\bibitem{tolman_static_1939}
R.C.~Tolman, \emph{Static {Solutions} of {Einstein}'s {Field} {Equations} for
  {Spheres} of {Fluid}},
  \href{https://doi.org/10.1103/PhysRev.55.364}{\emph{Phys. Rev.} {\bfseries
  55} (1939) 364}.

\bibitem{oppenheimer_massive_1939}
J.R.~Oppenheimer and G.M.~Volkoff, \emph{On {Massive} {Neutron} {Cores}},
  \href{https://doi.org/10.1103/PhysRev.55.374}{\emph{Phys. Rev.} {\bfseries
  55} (1939) 374}.

\bibitem{cowling_non-radial_1941}
T.G.~Cowling, \emph{The {Non}-radial {Oscillations} of {Polytropic} {Stars}},
  \href{https://doi.org/10.1093/mnras/101.8.367}{\emph{Monthly Notices of the
  Royal Astronomical Society} {\bfseries 101} (1941) 367}.

\bibitem{sotani_signatures_2011}
H.~Sotani, N.~Yasutake, T.~Maruyama and T.~Tatsumi, \emph{Signatures of
  hadron-quark mixed phase in gravitational waves},
  \href{https://doi.org/10.1103/PhysRevD.83.024014}{\emph{Phys. Rev. D}
  {\bfseries 83} (2011) 024014}.

\bibitem{das_impacts_2021}
H.~Das, A.~Kumar, S.~Biswal and S.~Patra, \emph{Impacts of dark matter on the f
  -mode oscillation of hyperon star},
  \href{https://doi.org/10.1103/PhysRevD.104.123006}{\emph{Phys. Rev. D}
  {\bfseries 104} (2021) 123006}.

\bibitem{panotopoulos_radial_2020}
G.~Panotopoulos, {\'A}.~Rinc{\'o}n and I.~Lopes, \emph{Radial oscillations and
  tidal {Love} numbers of dark energy stars},
  \href{https://doi.org/10.1140/epjp/s13360-020-00867-x}{\emph{Eur. Phys. J.
  Plus} {\bfseries 135} (2020) 856}.

\bibitem{rather_radial_2023}
I.A.~Rather, K.D.~Marquez, G.~Panotopoulos and I.~Lopes, \emph{Radial
  {Oscillations} in {Neutron} {Stars} with {Delta} {Baryons}},
  \href{https://doi.org/10.1103/PhysRevD.107.123022}{\emph{Phys. Rev. D}
  {\bfseries 107} (2023) 123022}.

\bibitem{routaray_radial_2023}
P.~Routaray, H.C.~Das, S.~Sen, B.~Kumar, G.~Panotopoulos and T.~Zhao,
  \emph{Radial oscillations of dark matter admixed neutron stars},
  \href{https://doi.org/10.1103/PhysRevD.107.103039}{\emph{Phys. Rev. D}
  {\bfseries 107} (2023) 103039}.

\bibitem{sen_radial_2023}
S.~Sen, S.~Kumar, A.~Kunjipurayil, P.~Routaray, S.~Ghosh, P.J.~Kalita et~al.,
  \emph{Radial {Oscillations} in {Neutron} {Stars} from {Unified} {Hadronic}
  and {Quarkyonic} {Equation} of {States}},
  \href{https://doi.org/10.3390/galaxies11020060}{\emph{Galaxies} {\bfseries
  11} (2023) 60}.

\bibitem{pinku_gw-posterior}
P.~Routaray, A.~Quddus, K.~Chakravarti and B.~Kumar, \emph{{Probing the impact
  of WIMP dark matter on universal relations, GW170817 posterior, and radial
  oscillations}}, \href{https://doi.org/10.1093/mnras/stad2628}{\emph{Monthly
  Notices of the Royal Astronomical Society} {\bfseries 525} (2023) 5492}.

\bibitem{mohanty_impact_2023}
S.R.~Mohanty, S.~Ghosh, P.~Routaray, H.C.~Das and B.~Kumar, \emph{The {Impact}
  of {Anisotropy} on {Neutron} {Star} {Properties}: {Insights} from {I}-f-{C}
  {Universal} {Relations}},  May, 2023.
\newblock 10.48550/arXiv.2305.15724.

\bibitem{bikram_fmode_2021}
B.K.~Pradhan and D.~Chatterjee, \emph{Effect of hyperons on $f$-mode
  oscillations in neutron stars},
  \href{https://doi.org/10.1103/PhysRevC.103.035810}{\emph{Phys. Rev. C}
  {\bfseries 103} (2021) 035810}.

\bibitem{bikram_GRfmode_2021}
B.K.~Pradhan, D.~Chatterjee, M.~Lanoye and P.~Jaikumar, \emph{General
  relativistic treatment of $f$-mode oscillations of hyperonic stars},
  \href{https://doi.org/10.1103/PhysRevC.106.015805}{\emph{Phys. Rev. C}
  {\bfseries 106} (2022) 015805}.

\bibitem{athul-fmode_2022}
A.~Kunjipurayil, T.~Zhao, B.~Kumar, B.K.~Agrawal and M.~Prakash, \emph{Impact
  of the equation of state on $f$- and $p$- mode oscillations of neutron
  stars}, \href{https://doi.org/10.1103/PhysRevD.106.063005}{\emph{Phys. Rev.
  D} {\bfseries 106} (2022) 063005}.

\bibitem{routaray_investigating_2023}
P.~Routaray, S.R.~Mohanty, H.~Das, S.~Ghosh, P.~Kalita, V.~Parmar et~al.,
  \emph{Investigating dark matter-admixed neutron stars with nitr equation of
  state in light of psr j0952-0607},
  \href{https://doi.org/10.1088/1475-7516/2023/10/073}{\emph{Journal of
  Cosmology and Astroparticle Physics} {\bfseries 2023} (2023) 073}.

\bibitem{routaray_dark_2023}
P.~Routaray, H.C.~Das, J.A.~Pattnaik and B.~Kumar, \emph{Dark {Matter}
  {Admixed} {Neutron} {Star} in the light of {HESS} {J1731}-347 and {PSR}
  {J0952}-0607},  July, 2023.
\newblock 10.48550/arXiv.2307.12748.

\bibitem{gearheart_upper_2011}
M.~Gearheart, W.G.~Newton, J.~Hooker and B.-A.~Li, \emph{Upper limits on the
  observational effects of nuclear pasta in neutron stars: {Effects} of nuclear
  pasta in neutron stars},
  \href{https://doi.org/10.1111/j.1365-2966.2011.19628.x}{\emph{Monthly Notices
  of the Royal Astronomical Society} {\bfseries 418} (2011) 2343}.

\bibitem{sotani_probing_2012}
H.~Sotani, K.~Nakazato, K.~Iida and K.~Oyamatsu, \emph{Probing the {Equation}
  of {State} of {Nuclear} {Matter} via {Neutron} {Star} {Asteroseismology}},
  \href{https://doi.org/10.1103/PhysRevLett.108.201101}{\emph{Phys. Rev. Lett.}
  {\bfseries 108} (2012) 201101}.

\bibitem{sotani_effect_2013}
H.~Sotani, K.~Nakazato, K.~Iida and K.~Oyamatsu, \emph{Effect of superfluidity
  on neutron star oscillations},
  \href{https://doi.org/10.1093/mnrasl/sls006}{\emph{Monthly Notices of the
  Royal Astronomical Society: Letters} {\bfseries 428} (2013) L21}.

\bibitem{sotani_possible_2013}
H.~Sotani, K.~Nakazato, K.~Iida and K.~Oyamatsu, \emph{Possible constraints on
  the density dependence of the nuclear symmetry energy from quasi-periodic
  oscillations in soft gamma repeaters},
  \href{https://doi.org/10.1093/mnras/stt1152}{\emph{Monthly Notices of the
  Royal Astronomical Society} {\bfseries 434} (2013) 2060}.

\bibitem{sotani_possible_2016}
H.~Sotani, K.~Iida and K.~Oyamatsu, \emph{Possible identifications of newly
  observed magnetar quasi-periodic oscillations as crustal shear modes},
  \href{https://doi.org/10.1016/j.newast.2015.08.003}{\emph{New Astronomy}
  {\bfseries 43} (2016) 80}.

\bibitem{sotani_probing_2017}
H.~Sotani, K.~Iida and K.~Oyamatsu, \emph{Probing nuclear bubble structure via
  neutron star asteroseismology},
  \href{https://doi.org/10.1093/mnras/stw2575}{\emph{Mon. Not. R. Astron. Soc.}
  {\bfseries 464} (2017) 3101}.

\bibitem{sotani_constraints_2018}
H.~Sotani, K.~Iida and K.~Oyamatsu, \emph{Constraints on the nuclear equation
  of state and the neutron star structure from crustal torsional oscillations},
  \href{https://doi.org/10.1093/mnras/sty1755}{\emph{Monthly Notices of the
  Royal Astronomical Society} {\bfseries 479} (2018) 4735}.

\bibitem{sotani_astrophysical_2019}
H.~Sotani, K.~Iida and K.~Oyamatsu, \emph{Astrophysical implications of
  double-layer torsional oscillations in a neutron star crust as a lasagna
  sandwich}, \href{https://doi.org/10.1093/mnras/stz2385}{\emph{Monthly Notices
  of the Royal Astronomical Society} (2019) stz2385}.

\bibitem{andersson_gravitational_1996}
N.~Andersson and K.D.~Kokkotas, \emph{Gravitational {Waves} and {Pulsating}
  {Stars}: {What} {Can} {We} {Learn} from {Future} {Observations}?},
  \href{https://doi.org/10.1103/PhysRevLett.77.4134}{\emph{Phys. Rev. Lett.}
  {\bfseries 77} (1996) 4134}.

\bibitem{andersson_towards_1998}
N.~Andersson and K.D.~Kokkotas, \emph{Towards gravitational wave
  asteroseismology},
  \href{https://doi.org/10.1046/j.1365-8711.1998.01840.x}{\emph{Monthly Notices
  of the Royal Astronomical Society} {\bfseries 299} (1998) 1059}.

\bibitem{sotani_density_2001}
H.~Sotani, K.~Tominaga and K.-i.~Maeda, \emph{Density discontinuity of a
  neutron star and gravitational waves},
  \href{https://doi.org/10.1103/PhysRevD.65.024010}{\emph{Phys. Rev. D}
  {\bfseries 65} (2001) 024010}.

\bibitem{passamonti_towards_2012}
A.~Passamonti and N.~Andersson, \emph{Towards real neutron star seismology:
  accounting for elasticity and superfluidity},
  \href{https://doi.org/10.1111/j.1365-2966.2011.19725.x}{\emph{Monthly Notices
  of the Royal Astronomical Society} {\bfseries 419} (2012) 638}.

\bibitem{doneva_gravitational_2013}
D.D.~Doneva, E.~Gaertig, K.D.~Kokkotas and C.~Krüger, \emph{Gravitational wave
  asteroseismology of fast rotating neutron stars with realistic equations of
  state}, \href{https://doi.org/10.1103/PhysRevD.88.044052}{\emph{Phys. Rev. D}
  {\bfseries 88} (2013) 044052}.

\bibitem{sotani_gravitational_2020}
H.~Sotani, \emph{Gravitational wave asteroseismology for low-mass neutron
  stars}, \href{https://doi.org/10.1103/PhysRevD.102.063023}{\emph{Phys. Rev.
  D} {\bfseries 102} (2020) 063023}.

\bibitem{sotani_estimating_2020}
H.~Sotani, \emph{Estimating the nuclear saturation parameter via low-mass
  neutron star asteroseismology},
  \href{https://doi.org/10.1103/PhysRevD.102.103021}{\emph{Phys. Rev. D}
  {\bfseries 102} (2020) 103021}.

\bibitem{landau1932theory}
L.D.~Landau, \emph{On the theory of stars}, {\emph{Phys. Z. Sowjetunion}
  {\bfseries 1} (1932) 152}.

\bibitem{baade_cosmic_1934}
W.~Baade and F.~Zwicky, \emph{Cosmic {Rays} from {Super}-{Novae}},
  \href{https://doi.org/10.1073/pnas.20.5.259}{\emph{Proceedings of the
  National Academy of Sciences} {\bfseries 20} (1934) 259}.

\bibitem{glendenning_hyperon_1982}
N.K.~Glendenning, \emph{The hyperon composition of neutron stars},
  \href{https://doi.org/10.1016/0370-2693(82)90078-8}{\emph{Physics Letters B}
  {\bfseries 114} (1982) 392}.

\bibitem{prakash_quark-hadron_1995}
M.~Prakash, J.R.~Cooke and J.M.~Lattimer, \emph{Quark-hadron phase transition
  in protoneutron stars},
  \href{https://doi.org/10.1103/PhysRevD.52.661}{\emph{Phys. Rev. D} {\bfseries
  52} (1995) 661}.

\bibitem{baldo_hyperon_2000}
M.~Baldo, G.F.~Burgio and H.-J.~Schulze, \emph{Hyperon stars in the
  {Brueckner}-{Bethe}-{Goldstone} theory},
  \href{https://doi.org/10.1103/PhysRevC.61.055801}{\emph{Phys. Rev. C}
  {\bfseries 61} (2000) 055801}.

\bibitem{oertel_hyperons_2015}
M.~Oertel, C.~Providência, F.~Gulminelli and A.R.~Raduta, \emph{Hyperons in
  neutron star matter within relativistic mean-field models},
  \href{https://doi.org/10.1088/0954-3899/42/7/075202}{\emph{J. Phys. G: Nucl.
  Part. Phys.} {\bfseries 42} (2015) 075202}.

\bibitem{vidana_hyperons_2016}
I.~Vidaña, \emph{Hyperons in {Neutron} {Stars}},
  \href{https://doi.org/10.1088/1742-6596/668/1/012031}{\emph{J. Phys.: Conf.
  Ser.} {\bfseries 668} (2016) 012031}.

\bibitem{marquez_phase_2017}
K.D.~Marquez and D.P.~Menezes, \emph{Phase transition in compact stars:
  nucleation mechanism and $\gamma$-ray bursts revisited},
  \href{https://doi.org/10.1088/1475-7516/2017/12/028}{\emph{J. Cosmol.
  Astropart. Phys.} {\bfseries 2017} (2017) 028}.

\bibitem{roark_hyperons_2019}
J.~Roark, X.~Du, C.~Constantinou, V.~Dexheimer, A.W.~Steiner and J.R.~Stone,
  \emph{Hyperons and quarks in proto-neutron stars},
  \href{https://doi.org/10.1093/mnras/stz1240}{\emph{Monthly Notices of the
  Royal Astronomical Society} {\bfseries 486} (2019) 5441}.

\bibitem{stone_equation_2021}
J.R.~Stone, V.~Dexheimer, P.A.M.~Guichon, A.W.~Thomas and S.~Typel,
  \emph{Equation of state of hot dense hyperonic matter in the
  {Quark}–{Meson}-{Coupling} ({QMC}-{A}) model},
  \href{https://doi.org/10.1093/mnras/staa4006}{\emph{Monthly Notices of the
  Royal Astronomical Society} {\bfseries 502} (2021) 3476}.

\bibitem{sedrakian_confronting_2020}
A.~Sedrakian, F.~Weber and J.J.~Li, \emph{Confronting {GW190814} with
  hyperonization in dense matter and hypernuclear compact stars},
  \href{https://doi.org/10.1103/PhysRevD.102.041301}{\emph{Phys. Rev. D}
  {\bfseries 102} (2020) 041301}.

\bibitem{menezes_neutron_2021}
D.P.~Menezes, \emph{A {Neutron} {Star} {Is} {Born}},
  \href{https://doi.org/10.3390/universe7080267}{\emph{Universe} {\bfseries 7}
  (2021) 267}.

\bibitem{motta_role_2022}
T.F.~Motta and A.W.~Thomas, \emph{The role of baryon structure in neutron
  stars}, \href{https://doi.org/10.1142/S0217732322300014}{\emph{Mod. Phys.
  Lett. A} {\bfseries 37} (2022) 2230001}.

\bibitem{marquez_delta_2022}
K.D.~Marquez, D.P.~Menezes, H.~Pais and C.~Provid\^{e}ncia,
  \emph{$\mathrm{\Delta}$ baryons in neutron stars},
  \href{https://doi.org/10.1103/PhysRevC.106.055801}{\emph{Phys. Rev. C}
  {\bfseries 106} (2022) 055801}.

\bibitem{issifu_exotic_2023}
A.~Issifu, K.D.~Marquez, M.R.~Pelicer and D.P.~Menezes, \emph{Exotic baryons in
  hot neutron stars},
  \href{https://doi.org/10.1093/mnras/stad1198}{\emph{Monthly Notices of the
  Royal Astronomical Society} {\bfseries 522} (2023) 3263}.

\bibitem{li_implications_2019}
J.J.~Li and A.~Sedrakian, \emph{Implications from {GW170817} for
  $\mathrm{\Delta}$-isobar {Admixed} {Hypernuclear} {Compact} {Stars}},
  \href{https://doi.org/10.3847/2041-8213/ab1090}{\emph{ApJL} {\bfseries 874}
  (2019) L22}.

\bibitem{malfatti_delta_2020}
G.~Malfatti, M.G.~Orsaria, I.F.~Ranea-Sandoval, G.A.~Contrera and F.~Weber,
  \emph{Delta baryons and diquark formation in the cores of neutron stars},
  \href{https://doi.org/10.1103/PhysRevD.102.063008}{\emph{Phys. Rev. D}
  {\bfseries 102} (2020) 063008}.

\bibitem{li_rapidly_2020}
J.J.~Li, A.~Sedrakian and F.~Weber, \emph{Rapidly rotating
  $\mathrm{\Delta}$-resonance-admixed hypernuclear compact stars},
  \href{https://doi.org/10.1016/j.physletb.2020.135812}{\emph{Physics Letters
  B} {\bfseries 810} (2020) 135812}.

\bibitem{thapa_equation_2020}
V.B.~Thapa, M.~Sinha, J.-J.~Li and A.~Sedrakian, \emph{Equation of state of
  strongly magnetized matter with hyperons and $\mathrm{\Delta}$-resonances},
  \href{https://doi.org/10.3390/particles3040043}{\emph{Particles} {\bfseries
  3} (2020) 660}.

\bibitem{backes_effects_2021}
B.C.T.~Backes, K.D.~Marquezb and D.P.~Menezes, \emph{Effects of strong magnetic
  fields on the hadron-quark deconfinement transition},
  \href{https://doi.org/10.1140/epja/s10050-021-00544-2}{\emph{Eur. Phys. J. A}
  {\bfseries 57} (2021) 229}.

\bibitem{thapa_massive_2021}
V.B.~Thapa, M.~Sinha, J.J.~Li and A.~Sedrakian, \emph{Massive
  $\mathrm{\Delta}$-resonance admixed hypernuclear stars with anti-kaon
  condensations},
  \href{https://doi.org/10.1103/PhysRevD.103.063004}{\emph{Phys. Rev. D}
  {\bfseries 103} (2021) 063004}.

\bibitem{dexheimer_delta_2021}
V.~Dexheimer, K.D.~Marquez and D.P.~Menezes, \emph{Delta {Baryons} in
  {Neutron}-{Star} {Matter} under {Strong} {Magnetic} {Fields}},
  \href{https://doi.org/10.1140/epja/s10050-021-00532-6}{\emph{Eur. Phys. J. A}
  {\bfseries 57} (2021) 216}.

\bibitem{raduta_equations_2022}
A.R.~Raduta, \emph{Equations of state for hot neutron stars-{II}. {The} role of
  exotic particle degrees of freedom},
  \href{https://doi.org/10.1140/epja/s10050-022-00772-0}{\emph{Eur. Phys. J. A}
  {\bfseries 58} (2022) 115}.

\bibitem{marczenko_chiral_2022}
M.~Marczenko, K.~Redlich and C.~Sasaki, \emph{Chiral symmetry restoration and
  $\mathrm{\Delta}$ matter formation in neutron stars},
  \href{https://doi.org/10.1103/PhysRevD.105.103009}{\emph{Phys. Rev. D}
  {\bfseries 105} (2022) 103009}.

\bibitem{demorest_2010}
P.B.~Demorest, T.~Pennucci, S.M.~Ransom, M.S.E.~Roberts and J.W.T.~Hessels,
  \emph{A two-solar-mass neutron star measured using shapiro delay},
  \href{https://doi.org/10.1038/nature09466}{\emph{Nature} {\bfseries 467}
  (2010) 1081}.

\bibitem{arzoumanian_nanograv_2018}
Z.~Arzoumanian, A.~Brazier, S.~Burke-Spolaor, S.~Chamberlin, S.~Chatterjee,
  B.~Christy et~al., \emph{The nanograv 11-year data set: High-precision timing
  of 45 millisecond pulsars},
  \href{https://doi.org/10.3847/1538-4365/aab5b0}{\emph{ApJS} {\bfseries 235}
  (2018) 37}.

\bibitem{fonseca_nanograv_2016}
E.~Fonseca, T.T.~Pennucci, J.A.~Ellis, I.H.~Stairs, D.J.~Nice, S.M.~Ransom
  et~al., \emph{{THE} {NANOGRAV} {NINE}-{YEAR} {DATA} {SET}: {MASS} {AND}
  {GEOMETRIC} {MEASUREMENTS} {OF} {BINARY} {MILLISECOND} {PULSARS}},
  \href{https://doi.org/10.3847/0004-637X/832/2/167}{\emph{ApJ} {\bfseries 832}
  (2016) 167}.

\bibitem{ozel_massive_2010}
F.~Özel, D.~Psaltis, S.~Ransom, P.~Demorest and M.~Alford, \emph{The massive
  pulsar psr j1614–2230: Linking quantum chromodynamics, gamma-ray bursts,
  and gravitational wave astronomy},
  \href{https://doi.org/10.1088/2041-8205/724/2/L199}{\emph{ApJ} {\bfseries
  724} (2010) L199}.

\bibitem{riley_nicer_2021}
T.E.~Riley, A.L.~Watts, P.S.~Ray, S.~Bogdanov, S.~Guillot, S.M.~Morsink et~al.,
  \emph{A {NICER} {View} of the {Massive} {Pulsar} {PSR} {J0740}+6620
  {Informed} by {Radio} {Timing} and {XMM}-{Newton} {Spectroscopy}},
  \href{https://doi.org/10.3847/2041-8213/ac0a81}{\emph{ApJL} {\bfseries 918}
  (2021) L27}.

\bibitem{fonseca_refined_2021}
E.~Fonseca, H.T.~Cromartie, T.T.~Pennucci, P.S.~Ray, A.Y.~Kirichenko,
  S.M.~Ransom et~al., \emph{Refined {Mass} and {Geometric} {Measurements} of
  the {High}-mass {PSR} {J0740}+6620},
  \href{https://doi.org/10.3847/2041-8213/ac03b8}{\emph{ApJL} {\bfseries 915}
  (2021) L12}.

\bibitem{riley_nicer_2019}
T.E.~Riley, A.L.~Watts, S.~Bogdanov, P.S.~Ray, R.M.~Ludlam, S.~Guillot et~al.,
  \emph{A \textit{{NICER}} {View} of {PSR} {J0030}+0451: {Millisecond} {Pulsar}
  {Parameter} {Estimation}},
  \href{https://doi.org/10.3847/2041-8213/ab481c}{\emph{ApJ} {\bfseries 887}
  (2019) L21}.

\bibitem{miller_psr_2019}
M.C.~Miller, F.K.~Lamb, A.J.~Dittmann, S.~Bogdanov, Z.~Arzoumanian,
  K.C.~Gendreau et~al., \emph{{PSR} {J0030}+0451 {Mass} and {Radius} from
  \textit{{NICER}} {Data} and {Implications} for the {Properties} of {Neutron}
  {Star} {Matter}}, \href{https://doi.org/10.3847/2041-8213/ab50c5}{\emph{ApJ}
  {\bfseries 887} (2019) L24}.

\bibitem{abbott_gw190814_2020}
R.~Abbott, T.~Abbott, S.~Abraham, F.~Acernese, K.~Ackley, C.~Adams et~al.,
  \emph{Gw190814: gravitational waves from the coalescence of a 23 solar mass
  black hole with a 2.6 solar mass compact object},
  \href{https://doi.org/10.3847/2041-8213/ab960f}{\emph{ApJL} {\bfseries 896}
  (2020) L44}.

\bibitem{abbott_gw170817_2017}
B.P.~Abbott, R.~Abbott, T.D.~Abbott, F.~Acernese, K.~Ackley, {LIGO Scientific
  Collaboration and Virgo Collaboration} et~al., \emph{Gw170817: Observation of
  gravitational waves from a binary neutron star inspiral},
  \href{https://doi.org/10.1103/PhysRevLett.119.161101}{\emph{Phys. Rev. Lett.}
  {\bfseries 119} (2017) 161101}.

\bibitem{ligo_virgo_properties_2019}
{LIGO Scientific Collaboration and Virgo Collaboration}, B.P.~Abbott,
  R.~Abbott, T.D.~Abbott and {others}, \emph{Properties of the {Binary}
  {Neutron} {Star} {Merger} {GW170817}},
  \href{https://doi.org/10.1103/PhysRevX.9.011001}{\emph{Phys. Rev. X}
  {\bfseries 9} (2019) 011001}.

\bibitem{schaffner_hyperon-rich_1996}
J.~Schaffner and I.N.~Mishustin, \emph{Hyperon-rich matter in neutron stars},
  \href{https://doi.org/10.1103/PhysRevC.53.1416}{\emph{Phys. Rev. C}
  {\bfseries 53} (1996) 1416}.

\bibitem{wu_strange_2011}
C.~Wu and Z.~Ren, \emph{Strange hadronic stars in relativistic mean-field
  theory with the fsugold parameter set},
  \href{https://doi.org/10.1103/PhysRevC.83.025805}{\emph{Phys. Rev. C}
  {\bfseries 83} (2011) 025805}.

\bibitem{sun_strangeness_2019}
T.-T.~Sun, S.-S.~Zhang, Q.-L.~Zhang and C.-J.~Xia, \emph{Strangeness and
  $\mathrm{\Delta}$ resonance in compact stars with relativistic-mean-field
  models}, \href{https://doi.org/10.1103/PhysRevD.99.023004}{\emph{Phys. Rev.
  D} {\bfseries 99} (2019) 023004}.

\bibitem{maslov_hyperons_2017}
K.A.~Maslov, E.E.~Kolomeitsev and D.N.~Voskresensky, \emph{Hyperons,
  $\mathrm{\Delta}$ resonances and condensate of charged $\rho$ mesons within
  relativistic mean-field models with scaled hadron masses and couplings},
  \href{https://doi.org/10.1088/1742-6596/941/1/012053}{\emph{J. Phys.: Conf.
  Ser.} {\bfseries 941} (2017) 012053}.

\bibitem{kolomeitsev_delta_2017}
E.E.~Kolomeitsev, K.A.~Maslov and D.N.~Voskresensky, \emph{Delta isobars in
  relativistic mean-field models with $\sigma$-scaled hadron masses and
  couplings},
  \href{https://doi.org/10.1016/j.nuclphysa.2017.02.004}{\emph{Nuclear Physics
  A} {\bfseries 961} (2017) 106}.

\bibitem{sedrakian_delta-resonances_2022}
A.~Sedrakian and A.~Harutyunyan, \emph{Delta-resonances and hyperons in
  proto-neutron stars and merger remnants},
  \href{https://doi.org/10.1140/epja/s10050-022-00792-w}{\emph{Eur. Phys. J. A}
  {\bfseries 58} (2022) 137}.

\bibitem{drago_early_2014}
A.~Drago, A.~Lavagno, G.~Pagliara and D.~Pigato, \emph{Early appearance of
  $\mathrm{\Delta}$ isobars in neutron stars},
  \href{https://doi.org/10.1103/PhysRevC.90.065809}{\emph{Phys. Rev. C}
  {\bfseries 90} (2014) 065809}.

\bibitem{glendenning_neutron_1985}
N.K.~Glendenning, \emph{Neutron stars are giant hypernuclei?},
  \href{https://doi.org/10.1086/163253}{\emph{ApJ} {\bfseries 293} (1985) 470}.

\bibitem{schurhoff_neutron_2010}
T.~Schürhoff, S.~Schramm and V.~Dexheimer, \emph{{NEUTRON} {STARS} {WITH}
  {SMALL} {RADII}—{THE} {ROLE} {OF} $\mathrm{\Delta}$ {RESONANCES}},
  \href{https://doi.org/10.1088/2041-8205/724/1/L74}{\emph{ApJL} {\bfseries
  724} (2010) L74}.

\bibitem{thapa_dense_2020}
V.B.~Thapa and M.~Sinha, \emph{Dense matter equation of state of a massive
  neutron star with antikaon condensation},
  \href{https://doi.org/10.1103/PhysRevD.102.123007}{\emph{Phys. Rev. D}
  {\bfseries 102} (2020) 123007}.

\bibitem{maruyama_finite_2006}
T.~Maruyama, T.~Tatsumi, D.N.~Voskresensky, T.~Tanigawa, T.~Endo and S.~Chiba,
  \emph{Finite size effects on kaonic “pasta” structures},
  \href{https://doi.org/10.1103/PhysRevC.73.035802}{\emph{Phys. Rev. C}
  {\bfseries 73} (2006) 035802}.

\bibitem{brown_strangeness_2006}
G.E.~Brown, C.-H.~Lee, H.-J.~Park and M.~Rho, \emph{Strangeness condensation by
  expanding about the fixed point of the harada-yamawaki vector manifestation},
  \href{https://doi.org/10.1103/PhysRevLett.96.062303}{\emph{Phys. Rev. Lett.}
  {\bfseries 96} (2006) 062303}.

\bibitem{shao_influence_2010}
G.-y.~Shao and Y.-x.~Liu, \emph{Influence of the isovector-scalar channel
  interaction on neutron star matter with hyperons and antikaon condensation},
  \href{https://doi.org/10.1103/PhysRevC.82.055801}{\emph{Phys. Rev. C}
  {\bfseries 82} (2010) 055801}.

\bibitem{char_massive_2014}
P.~Char and S.~Banik, \emph{Massive neutron stars with antikaon condensates in
  a density-dependent hadron field theory},
  \href{https://doi.org/10.1103/PhysRevC.90.015801}{\emph{Phys. Rev. C}
  {\bfseries 90} (2014) 015801}.

\bibitem{burgio_hyperon_2011}
G.F.~Burgio, H.-J.~Schulze and A.~Li, \emph{Hyperon stars at finite temperature
  in the {Brueckner} theory},
  \href{https://doi.org/10.1103/PhysRevC.83.025804}{\emph{Phys. Rev. C}
  {\bfseries 83} (2011) 025804}.

\bibitem{lonardoni_hyperon_2015}
D.~Lonardoni, A.~Lovato, S.~Gandolfi and F.~Pederiva, \emph{Hyperon {Puzzle}:
  {Hints} from {Quantum} {Monte} {Carlo} {Calculations}},
  \href{https://doi.org/10.1103/PhysRevLett.114.092301}{\emph{Phys. Rev. Lett.}
  {\bfseries 114} (2015) 092301}.

\bibitem{bombaci_hyperon_2017}
I.~Bombaci, \emph{The {Hyperon} {Puzzle} in {Neutron} {Stars}},  in
  \emph{Proceedings of the 12th {International} {Conference} on {Hypernuclear}
  and {Strange} {Particle} {Physics} ({HYP2015})}, vol.~17 of \emph{{JPS}
  {Conference} {Proceedings}}, Journal of the Physical Society of Japan (2017),
  \href{https://doi.org/10.7566/JPSCP.17.101002}{DOI}.

\bibitem{drago_can_2014}
A.~Drago, A.~Lavagno and G.~Pagliara, \emph{Can very compact and very massive
  neutron stars both exist?},
  \href{https://doi.org/10.1103/PhysRevD.89.043014}{\emph{Phys. Rev. D}
  {\bfseries 89} (2014) 043014}.

\bibitem{raduta_-admixed_2021}
A.R.~Raduta, \emph{$\mathrm{\Delta}$-admixed neutron stars: {Spinodal}
  instabilities and {dUrca} processes},
  \href{https://doi.org/10.1016/j.physletb.2021.136070}{\emph{Physics Letters
  B} {\bfseries 814} (2021) 136070}.

\bibitem{takeda_catalysis_2018}
Y.~Takeda, Y.~Kim and M.~Harada, \emph{Catalysis of partial chiral symmetry
  restoration by {Delta} matter},
  \href{https://doi.org/10.1103/PhysRevC.97.065202}{\emph{Phys. Rev. C}
  {\bfseries 97} (2018) 065202}.

\bibitem{glendenning_reconciliation_1991}
N.K.~Glendenning and S.A.~Moszkowski, \emph{Reconciliation of neutron-star
  masses and binding of the $\mathrm{\Lambda}$ in hypernuclei},
  \href{https://doi.org/10.1103/PhysRevLett.67.2414}{\emph{Phys. Rev. Lett.}
  {\bfseries 67} (1991) 2414}.

\bibitem{li_competition_2018}
J.J.~Li, A.~Sedrakian and F.~Weber, \emph{Competition between delta isobars and
  hyperons and properties of compact stars},
  \href{https://doi.org/10.1016/j.physletb.2018.06.051}{\emph{Physics Letters
  B} {\bfseries 783} (2018) 234}.

\bibitem{ribes_interplay_2019}
P.~Ribes, A.~Ramos, L.~Tolos, C.~Gonzalez-Boquera and M.~Centelles,
  \emph{Interplay between {Delta} {Particles} and {Hyperons} in {Neutron}
  {Stars}}, \href{https://doi.org/10.3847/1538-4357/ab3a93}{\emph{ApJ}
  {\bfseries 883} (2019) 168}.

\bibitem{xiang_ensuremathdelta_2003}
H.~Xiang and G.~Hua, \emph{$\mathrm{\Delta}$ excitation and its influences on
  neutron stars in relativistic mean field theory},
  \href{https://doi.org/10.1103/PhysRevC.67.038801}{\emph{Phys. Rev. C}
  {\bfseries 67} (2003) 038801}.

\bibitem{lavagno_hot_2010}
A.~Lavagno, \emph{Hot and dense hadronic matter in an effective mean-field
  approach}, \href{https://doi.org/10.1103/PhysRevC.81.044909}{\emph{Phys. Rev.
  C} {\bfseries 81} (2010) 044909}.

\bibitem{cai_critical_2015}
B.-J.~Cai, F.J.~Fattoyev, B.-A.~Li and W.G.~Newton, \emph{Critical {Density}
  and {Impact} of $\mathrm{\Delta} (1232)$ {Resonance} {Formation} in {Neutron}
  {Stars}}, \href{https://doi.org/10.1103/PhysRevC.92.015802}{\emph{Phys. Rev.
  C} {\bfseries 92} (2015) 015802}.

\bibitem{bhuyan_attribute_2017}
M.~Bhuyan, B.V.~Carlson, S.K.~Patra and S.-G.~Zhou, \emph{The attribute of
  rotational profile to the hyperon puzzle in the prediction of heaviest
  compact star}, \href{https://doi.org/10.1142/S0218301317500525}{\emph{Int. J.
  Mod. Phys. E} {\bfseries 26} (2017) 1750052}.

\bibitem{weissenborn_hyperons_2012}
S.~Weissenborn, D.~Chatterjee and J.~Schaffner-Bielich, \emph{Hyperons and
  massive neutron stars: {Vector} repulsion and {SU}(3) symmetry},
  \href{https://doi.org/10.1103/PhysRevC.85.065802}{\emph{Phys. Rev. C}
  {\bfseries 85} (2012) 065802}.

\bibitem{ma_kaon_2022}
F.~Ma, W.~Guo and C.~Wu, \emph{Kaon meson condensate in neutron star matter
  including hyperons},
  \href{https://doi.org/10.1103/PhysRevC.105.015807}{\emph{Phys. Rev. C}
  {\bfseries 105} (2022) 015807}.

\bibitem{patra_effect_2022}
N.K.~Patra, B.K.~Sharma, A.~Reghunath, A.K.H.~Das and T.K.~Jha, \emph{Effect of
  the $\sigma$-cut potential on the properties of neutron stars with or without
  a hyperonic core},
  \href{https://doi.org/10.1103/PhysRevC.106.055806}{\emph{Phys. Rev. C}
  {\bfseries 106} (2022) 055806}.

\bibitem{maslov_making_2015}
K.A.~Maslov, E.E.~Kolomeitsev and D.N.~Voskresensky, \emph{Making a soft
  relativistic mean-field equation of state stiffer at high density},
  \href{https://doi.org/10.1103/PhysRevC.92.052801}{\emph{Phys. Rev. C}
  {\bfseries 92} (2015) 052801}.

\bibitem{zhang_massive_2018}
Y.~Zhang, J.~Hu and P.~Liu, \emph{Massive neutron star with strangeness in a
  relativistic mean-field model with a high-density cutoff},
  \href{https://doi.org/10.1103/PhysRevC.97.015805}{\emph{Phys. Rev. C}
  {\bfseries 97} (2018) 015805}.

\bibitem{malik_equation--state_2021}
T.~Malik, S.~Banik and D.~Bandyopadhyay, \emph{Equation-of-state {Table} with
  {Hyperon} and {Antikaon} for {Supernova} and {Neutron} {Star} {Merger}},
  \href{https://doi.org/10.3847/1538-4357/abe860}{\emph{ApJ} {\bfseries 910}
  (2021) 96}.

\bibitem{sun_neutron_2008}
B.Y.~Sun, W.H.~Long, J.~Meng and U.~Lombardo, \emph{Neutron star properties in
  density-dependent relativistic {Hartree}-{Fock} theory},
  \href{https://doi.org/10.1103/PhysRevC.78.065805}{\emph{Phys. Rev. C}
  {\bfseries 78} (2008) 065805}.

\bibitem{banik_density_2002}
S.~Banik and D.~Bandyopadhyay, \emph{Density dependent hadron field theory for
  neutron stars with antikaon condensates},
  \href{https://doi.org/10.1103/PhysRevC.66.065801}{\emph{Phys. Rev. C}
  {\bfseries 66} (2002) 065801}.

\bibitem{char_generalised_2023}
P.~Char, C.~Mondal, F.~Gulminelli and M.~Oertel, \emph{Generalised description
  of {Neutron} {Star} matter with nucleonic {Relativistic} {Density}
  {Functional}},  July, 2023.
\newblock 10.48550/arXiv.2307.12364.

\bibitem{colucci_equation_2013}
G.~Colucci and A.~Sedrakian, \emph{Equation of state of hypernuclear matter:
  {Impact} of hyperon--scalar-meson couplings},
  \href{https://doi.org/10.1103/PhysRevC.87.055806}{\emph{Phys. Rev. C}
  {\bfseries 87} (2013) 055806}.

\bibitem{taninah_parametric_2020}
A.~Taninah, S.E.~Agbemava, A.V.~Afanasjev and P.~Ring, \emph{Parametric
  correlations in energy density functionals},
  \href{https://doi.org/10.1016/j.physletb.2019.135065}{\emph{Physics Letters
  B} {\bfseries 800} (2020) 135065}.

\bibitem{kokkotas_radial_2001}
K.D.~Kokkotas and J.~Ruoff, \emph{Radial oscillations of relativistic stars},
  \href{https://doi.org/10.1051/0004-6361:20000216}{\emph{A\&A} {\bfseries 366}
  (2001) 565}.

\bibitem{li_oscillation_2022}
H.-B.~Li, Y.~Gao, L.~Shao, R.-X.~Xu and R.~Xu, \emph{Oscillation modes and
  gravitational waves from strangeon stars},
  \href{https://doi.org/10.1093/mnras/stac2622}{\emph{Mon Not R Astron Soc}
  {\bfseries 516} (2022) 6172}.

\bibitem{sagun_asteroseismology_2020}
V.~Sagun, G.~Panotopoulos and I.~Lopes, \emph{Asteroseismology: {Radial}
  oscillations of neutron stars with realistic equation of state},
  \href{https://doi.org/10.1103/PhysRevD.101.063025}{\emph{Phys. Rev. D}
  {\bfseries 101} (2020) 063025}.

\bibitem{panotopoulos_radial_2018}
G.~Panotopoulos and I.~Lopes, \emph{Radial oscillations of strange quark stars
  admixed with fermionic dark matter},
  \href{https://doi.org/10.1103/PhysRevD.98.083001}{\emph{Phys. Rev. D}
  {\bfseries 98} (2018) 083001}.

\bibitem{chen_building_2014}
W.-C.~Chen and J.~Piekarewicz, \emph{Building relativistic mean field models
  for finite nuclei and neutron stars},
  \href{https://doi.org/10.1103/PhysRevC.90.044305}{\emph{Phys. Rev. C}
  {\bfseries 90} (2014) 044305}.

\bibitem{lavagno_strangeness_2022}
A.~Lavagno and D.~Pigato, \emph{Strangeness thermodynamic instabilities in hot
  and dense nuclear matter},
  \href{https://doi.org/10.1140/epja/s10050-022-00885-6}{\emph{Eur. Phys. J. A}
  {\bfseries 58} (2022) 237}.

\bibitem{paoli_2013}
M.G.d.~Paoli, L.B.~Castro, D.P.~Menezes and C.C.~Barros,
  \emph{Rarita–schwinger particles under the influence of strong magnetic
  fields}, \href{https://doi.org/10.1088/0954-3899/40/5/055007}{\emph{J. Phys.
  G: Nucl. Part. Phys.} {\bfseries 40} (2013) 055007}.

\bibitem{hofmann_2001}
F.~Hofmann, C.M.~Keil and H.~Lenske, \emph{Density dependent hadron field
  theory for asymmetric nuclear matter and exotic nuclei},
  \href{https://doi.org/10.1103/PhysRevC.64.034314}{\emph{Phys. Rev. C}
  {\bfseries 64} (2001) 034314}.

\bibitem{schaffner_multiply_1994}
J.~Schaffner, C.~Dover, A.~Gal, C.~Greiner, D.~Millener and H.~Stocker,
  \emph{Multiply {Strange} {Nuclear} {Systems}},
  \href{https://doi.org/10.1006/aphy.1994.1090}{\emph{Annals of Physics}
  {\bfseries 235} (1994) 35}.

\bibitem{dover_hyperon-nucleus_1984}
C.B.~Dover and A.~Gal, \emph{Hyperon-nucleus potentials},
  \href{https://doi.org/10.1016/0146-6410(84)90004-8}{\emph{Progress in
  Particle and Nuclear Physics} {\bfseries 12} (1984) 171}.

\bibitem{glendenning_compact_1996}
N.~Glendenning, \emph{Compact {Stars}. {Nuclear} {Physics}, {Particle}
  {Physics} and {General} {Relativity}.} (Jan., 1996).

\bibitem{pais_dynamical_1966}
A.~Pais, \emph{Dynamical {Symmetry} in {Particle} {Physics}},
  \href{https://doi.org/10.1103/RevModPhys.38.215}{\emph{Rev. Mod. Phys.}
  {\bfseries 38} (1966) 215}.

\bibitem{lopes_hypernuclear_2014}
L.L.~Lopes and D.P.~Menezes, \emph{Hypernuclear matter in a complete {SU}(3)
  symmetry group},
  \href{https://doi.org/10.1103/PhysRevC.89.025805}{\emph{Phys. Rev. C}
  {\bfseries 89} (2014) 025805}.

\bibitem{jaiswal_constraining_2021}
S.~Jaiswal and D.~Chatterjee, \emph{Constraining {Dense} {Matter} {Physics}
  {Using} f-{Mode} {Oscillations} in {Neutron} {Stars}},
  \href{https://doi.org/10.3390/physics3020022}{\emph{Physics} {\bfseries 3}
  (2021) 302}.

\bibitem{douchin_unified_2001}
F.~Douchin and P.~Haensel, \emph{A unified equation of state of dense matter
  and neutron star structure},
  \href{https://doi.org/10.1051/0004-6361:20011402}{\emph{A\&A} {\bfseries 380}
  (2001) 151}.

\bibitem{hinderer_tidal_2008}
T.~Hinderer, \emph{Tidal {Love} {Numbers} of {Neutron} {Stars}},
  \href{https://doi.org/10.1086/533487}{\emph{ApJ} {\bfseries 677} (2008)
  1216}.

\bibitem{hinderer_tidal_2010}
T.~Hinderer, B.D.~Lackey, R.N.~Lang and J.S.~Read, \emph{Tidal deformability of
  neutron stars with realistic equations of state and their gravitational wave
  signatures in binary inspiral},
  \href{https://doi.org/10.1103/PhysRevD.81.123016}{\emph{Phys. Rev. D}
  {\bfseries 81} (2010) 123016}.

\bibitem{flanagan_constraining_2008}
{\'E}.{\'E}.~Flanagan and T.~Hinderer, \emph{Constraining neutron-star tidal
  {Love} numbers with gravitational-wave detectors},
  \href{https://doi.org/10.1103/PhysRevD.77.021502}{\emph{Phys. Rev. D}
  {\bfseries 77} (2008) 021502}.

\bibitem{vasquez_flores_2014}
C.~Vásquez~Flores and G.~Lugones, \emph{Discriminating hadronic and quark
  stars through gravitational waves of fluid pulsation modes},
  \href{https://doi.org/10.1088/0264-9381/31/15/155002}{\emph{Class. Quantum
  Grav.} {\bfseries 31} (2014) 155002}.

\bibitem{ranea-sandoval_2018}
I.F.~Ranea-Sandoval, O.M.~Guilera, M.~Mariani and M.G.~Orsaria,
  \emph{Oscillation modes of hybrid stars within the relativistic cowling
  approximation}, \href{https://doi.org/10.1088/1475-7516/2018/12/031}{\emph{J.
  Cosmol. Astropart. Phys.} {\bfseries 2018} (2018) 031}.

\end{thebibliography}\endgroup
\end{document}